\documentclass[paper]{JHEP3}

\usepackage{epsfig}
\usepackage{graphicx}
\usepackage{feynmp}

\usepackage{amsmath}
\usepackage{amsfonts}
\usepackage{amssymb}
\usepackage{graphicx}

\makeatletter
\def\@fpheader{\\}
\makeatother

\newif\ifdraft
\drafttrue
\newif\ifpreprint
\preprinttrue

\def\Sect#1{Section~{\ref{#1}}}
\def\sect#1{section~{\ref{#1}}}
\def\beq{\begin{equation}}
\def\eeq{\end{equation}}
\def\beqa{\begin{eqnarray}}
\def\eeqa{\end{eqnarray}}
\def\be{\begin{equation}}
\def\ee{\end{equation}}
\def\bea{\begin{eqnarray}}
\def\eea{\end{eqnarray}}
\newcommand{\nn}{\nonumber}
\def\bsp#1\esp{\begin{split}#1\end{split}}
\def\Eqn#1{Equation~(\ref{#1})}
\def\eqn#1{eq.~(\ref{#1})}
\def\eqns#1#2{eqs.~(\ref{#1}) and~(\ref{#2})}

\def\Fig#1{Figure~\ref{#1}}
\def\fig#1{fig.~\ref{#1}}

\def\tree{{\rm tree}}
\def\oneloop{{\rm 1\! -\! loop}}
\def\e{\epsilon}

\def\to{\rightarrow}
\def\lr{\leftrightarrow}
\def\Tr{\mathop{\rm Tr}\nolimits}

\def\e{\epsilon}

\def\cg{c_\Gamma}
\def\hf{\textstyle{1\over2}}
\def\Li{\mathop{\rm Li}\nolimits}

\def\sandp#1.#2.#3{%
\left\langle\smash{#1}{\vphantom1}^{-}\right|{#2}%
\left|\smash{#3}{\vphantom1}^{+}\right\rangle}
\def\sandpp#1.#2.#3{%
\left\langle\smash{#1}{\vphantom1}^{+}\right|{#2}%
\left|\smash{#3}{\vphantom1}^{+}\right\rangle}
\def\sandmm#1.#2.#3{%
\left\langle\smash{#1}{\vphantom1}^{-}\right|{#2}%
\left|\smash{#3}{\vphantom1}^{-}\right\rangle}
\def\spab#1.#2.#3{\sandmm#1.#2.#3}
\def\spba#1.#2.#3{\sandpp#1.#2.#3}
\def\spaa#1.#2.#3.#4{\sandmp#1.{#2#3}.#4}
\def\spbb#1.#2.#3.#4{\sandpm#1.{#2#3}.#4}
\def\spa#1.#2{\left\langle#1\,#2\right\rangle}
\def\spb#1.#2{\left[#1\,#2\right]}
\def\spash#1.#2{\vphantom{\hat K}\spa{\smash{#1}}.{\smash{#2}}}
\def\spbsh#1.#2{\vphantom{\hat K}\spb{\smash{#1}}.{\smash{#2}}}
\def\lor#1.#2{\left(#1\,#2\right)}
\def\sand#1.#2.#3{%
\left\langle\smash{#1}{\vphantom1}^{-}\right|{#2}%
\left|\smash{#3}{\vphantom1}^{-}\right\rangle}
\def\sandpp#1.#2.#3{%
\left\langle\smash{#1}{\vphantom1}^{+}\right|{#2}%
\left|\smash{#3}{\vphantom1}^{+}\right\rangle}
\def\sandpm#1.#2.#3{%
\left\langle\smash{#1}{\vphantom1}^{+}\right|{#2}%
\left|\smash{#3}{\vphantom1}^{-}\right\rangle}
\def\sandmp#1.#2.#3{%
\left\langle\smash{#1}{\vphantom1}^{-}\right|{#2}%
\left|\smash{#3}{\vphantom1}^{+}\right\rangle}
\def\sandmppm#1.#2.#3{%
\left\langle\smash{#1^{\mp}}\vphantom1\right|{#2}%
\left|\smash{#3^{\pm}}{\vphantom1}\right\rangle}
\def\sandpmpm#1.#2.#3{%
\left\langle\smash{#1^{\pm}}\vphantom1\right|{#2}%
\left|\smash{#3^{\pm}}{\vphantom1}\right\rangle}

\newbox\charbox
\newbox\slabox
\def\s#1{{      
        \setbox\charbox=\hbox{$#1$}
        \setbox\slabox=\hbox{$/$}
        \dimen\charbox=\ht\slabox
        \advance\dimen\charbox by -\dp\slabox
        \advance\dimen\charbox by -\ht\charbox
        \advance\dimen\charbox by \dp\charbox
        \divide\dimen\charbox by 2
        \raise-\dimen\charbox\hbox to \wd\charbox{\hss/\hss}
        \llap{$#1$} }}

\def\lsl{\not{\hbox{\kern-2.3pt $\ell$}}}
\def\Ksl{\not{\hbox{\kern-2.3pt $K$}}}
\def\ksl{\not{\hbox{\kern-2.3pt $k$}}}
\def\qsl{\not{\hbox{\kern-2.3pt $q$}}}
\def\esl{\not{\hbox{\kern-2.3pt $\pol$}}}
\def\ib{{\bar\imath}}
\def\jb{{\bar\jmath}}

\def\llongrightarrow{%
\relbar\mskip-0.5mu\joinrel\mskip-0.5mu\relbar
     \mskip-0.5mu\joinrel\longrightarrow}
\def\inlimit^#1{\buildrel#1\over\llongrightarrow}

\def\cA{{\cal A}}
\def\I{{\cal I}}
\def\pol{\varepsilon}
\def\qb{{\bar{q}}}
\def\Split{\mathop{\rm Split}\nolimits}
\def\Ord{{\cal O}}
\def\Ellin{\ell}

\newcommand{\Bmp}[1]{\langle #1\rangle}

\newcommand{\Kfmu}[1]{\tilde K^{\mu}_{#1}}

\newcommand{\Kfm}[1]{\tilde K^{-}_{#1}}

\title{A brief introduction to modern amplitude methods}

\author{Lance J. Dixon\\
SLAC National Accelerator Laboratory\\
Stanford University, Stanford CA 94309, USA\\
E-mail: \email{lance@slac.stanford.edu}}

\abstract{I provide a basic introduction to modern helicity amplitude methods,
including color organization, the spinor helicity formalism, and
factorization properties.  I also describe the BCFW (on-shell) recursion
relation at tree level, and explain how similar ideas --- unitarity
and on-shell methods --- work at the loop level.  These notes are
based on lectures delivered at the 2012 CERN Summer School and at
TASI 2013.}

\keywords{helicity amplitudes, factorization, unitarity}
\preprint{SLAC--PUB--15775}

\begin{document}

\section{Introduction}
\label{sec:intro}

Scattering amplitudes are at the heart of high energy physics.
They lie at the intersection between quantum field theory and
collider experiments.  Currently we are in the hadron collider era,
which began at the Tevatron and has now moved to the Large Hadron
Collider (LHC).  Hadron colliders are broadband machines capable of great
discoveries, such as the Higgs boson~\cite{HiggsDiscovery},
but there are also huge Standard Model backgrounds to many potential
signals.  If we are to discover new physics (besides the Higgs boson)
at the LHC, we will need to understand the old physics of the Standard
Model at an exquisitely precise level.  QCD dominates collisions at the
LHC, and the largest theoretical uncertainties for most processes are due
to our limited knowledge of higher order terms in perturbative QCD.

Many theorists have been working to improve this situation.  Some have
been computing the next-to-leading order (NLO) QCD corrections to
complex collider processes that were previously only known at leading
order (LO).  LO uncertainties are often of order one, while NLO uncertainties
can be in the 10--20\% range, depending on the process.
Others have been computing the next-to-next-to-leading order (NNLO)
corrections to benchmark processes that are only known at NLO;
most NNLO predictions have uncertainties in the range of 1--5\%,
allowing precise experimental measurements to be interpreted with
similar theoretical precision.  Higher-order computations have a
number of technical ingredients, but they all require loop amplitudes,
one-loop for NLO, and both one- and two-loop for NNLO, as well as
tree amplitudes of higher multiplicity.

The usual textbook methods for computing an unpolarized cross section involve
squaring the scattering amplitude at the beginning, then
summing analytically over the spins of external states, and transforming
the result into an expression that only involves momentum invariants
(Mandelstam variables) and masses.  For complex processes, this approach
is usually infeasible.  If there are $N$ Feynman diagrams for an amplitude,
then there are $N^2$ terms in the square of the amplitude.  It is much
better to calculate the $N$ terms in the amplitude, as a complex number,
and then compute the cross section by squaring that number.
This approach of directly computing the amplitude benefits greatly
from the fact that many amplitudes are much simpler than one might
expect from the number of Feynman diagrams contributing to them.

In order to compute the amplitude directly, one has to pick a basis
for the polarization states of the external particles.  At collider
energies, most of these particles are effectively massless:
the light quarks and gluons, photons, and the charged leptons and neutrinos
(decay products of $W$ and $Z$ bosons).  Massless fermions
have the property that their chirality and helicity coincide,
and their chirality is preserved by the gauge interactions.
Therefore the helicity basis is clearly an optimal one for massless
fermions, because many matrix elements (the helicity-flip ones) will
always vanish.

Around three decades ago, it was realized that the helicity basis
was extremely useful for massless vector bosons as well~\cite{SpinorHelicity}.
Many tree-level amplitudes were found to vanish in this basis as well
(which could be explained by a secret supersymmetry obeyed by
tree amplitudes~\cite{OldSWI,NewSWI}).  Also, the nonvanishing
amplitudes were found to possess a hierarchy of simplicity, depending on
how much they violated helicity ``conservation''.  For example,
a simple one-term expression for the tree amplitudes for scattering
an arbitrary number of gluons with maximal helicity violation
(MHV) was found by Parke and Taylor in 1986~\cite{ParkeTaylor}, and proven
recursively by Berends and Giele shortly thereafter~\cite{BGRecursion}.

As the first loop computations were performed for gluon scattering
in the helicity basis~\cite{BKComputation,FiveGluon}, it became
apparent that (relative) simplicity of amplitudes 
could extend to the loop level.
One way to maintain the simplicity is to use unitarity~\cite{Unitarity}
to determine loop amplitudes by using tree amplitudes as input.
These methods have been refined enormously over the years, and
automated in order to handle very complicated processes.
They now form an important part of the arsenal for theorists providing
NLO results for LHC experiments.  Many of the methods
are now being further refined and extended to the two-loop level, and within
a few years we may see a similar NNLO arsenal come to full fruition.

Besides QCD, unitarity-based methods have also found widespread
application to scattering amplitudes for more formal theories, such
as ${\cal N}=4$ super-Yang-Mills theory and ${\cal N}=8$ supergravity,
just to mention a couple of examples.  The more supersymmetry,
the greater the simplicity of the amplitudes, allowing analytical
results to be obtained for many multi-loop amplitudes (at least
before integrating over the loop momentum).  These results have
helped to expose new symmetries, which have in turn led to other
powerful methods for computing in these special theories.

The purpose of these lecture notes is to provide a brief and basic
introduction to modern amplitude methods.  They are intended for someone
who has taken a first course in quantum field theory, but who has never
studied these methods before.  For someone who wants to go on further and
perform research using such methods in either QCD or more formal areas,
these notes will be far from sufficient.  Fortunately, there are much 
more thorough reviews available.  In particular, methods for one-loop
QCD amplitudes have been reviewed in
refs.~\cite{OnShellReview,BrittoReview,ItaReview,EKMZReview}.
Also, a very recent and comprehensive article~\cite{EHReview} covers much
of the material covered here, plus a great deal more, particularly in the
direction of methods for multi-loop amplitudes in more formal theories.
There are also reviews of basic tree-level organization and
properties~\cite{MPReview,LDTASI95,LDIntroAmp} and of one-loop
unitarity~\cite{AnnRev}.  Other reviews emphasize
${\cal N}=4$ super-Yang-Mills
theory~\cite{AldayRoiban,DrummondReview}.

These notes are organized as follows.
In \sect{sec:color} we describe trace-based color decompositions for
QCD amplitudes. In \sect{sec:helicity} we review the spinor helicity
formalism, and apply it to the computation of some simple four- and
five-point tree amplitudes.  In \sect{sec:factorization}
we use these results to illustrate the universal soft and
collinear factorization of gauge theory amplitudes.  We also
introduce the Parke-Taylor amplitudes, and discuss the utility
of spinor variables for describing collinear limits and massless
three-point kinematics.   In \sect{sec:trees} we explain the BCFW
(on-shell) recursion relation for tree amplitudes, and apply it
to the Parke-Taylor amplitudes, as well as to a next-to-MHV example.
\Sect{sec:loops} discusses the application of generalized unitarity
to one-loop amplitudes, and in \sect{sec:concl} we conclude.


\section{Color decompositions}
\label{sec:color}

In this section we explain how to organize the color degrees
of freedom in QCD amplitudes, in order to separate out pieces that
have simpler analytic properties.  Those pieces have various names
in the literature, such as color-ordered amplitudes,
dual amplitudes, primitive amplitudes and partial amplitudes.
(There is a distinction between primitive
amplitudes and partial amplitudes at the loop level, but not at tree level,
at least not unless there are multiple fermion lines.)

The basic idea~\cite{TreeColor,TreeColor2,MPReview,LDTASI95}
is to reorganize the color degrees of freedom of QCD, in order
to eliminate the Lie algebra structure constants $f^{abc}$ found
in the Feynman rules, in favor of the generator matrices $T^a$
in the fundamental representation of $SU(N_c)$.  Although the gauge
group of QCD is $SU(3)$, it requires no extra effort to generalize it
to $SU(N_c)$, and one can often gain insight by making the dependence
on $N_c$ explicit.  Sometimes it is also advantageous (especially
computationally) to consider the limit of a large number of colors,
$N_c\to\infty$.

Gluons in an $SU(N_c)$ gauge theory carry an adjoint color index
$a=1,2,\ldots,N_c^2-1$,  while quarks and antiquarks carry an 
$N_c$ or $\overline{N}_c$ index, $i,\jb=1,2,\ldots,N_c$.  
The generators of $SU(N_c)$ in the fundamental representation are
traceless hermitian $N_c\times N_c$ matrices, $(T^a)_i^{~\jb}$.
For computing color-ordered helicity amplitudes, it's conventional
to normalize them according to $\Tr(T^aT^b) = \delta^{ab}$
in order to avoid a proliferation of $\sqrt{2}$'s in the
amplitudes.

Each Feynman diagram in QCD contains a factor of $(T^a)_i^{~\jb}$
for each gluon-quark-anti-quark vertex,
a group theory structure constant $f^{abc}$ for each pure gluon 
three-point vertex, and contracted pairs of structure constants
$f^{abe}f^{cde}$ for each pure gluon four-vertex. 
The structure constants are defined by the commutator
\be
[T^a,T^b]\ =\ i\sqrt{2}\,f^{abc}\,T^c \,.
\label{fabcdef}
\ee
The internal gluon and quark propagators contract indices
together with factors of $\delta_{ab}$, $\delta^{~\jb}_{i}$.
We want to identify all possible color factors for the diagrams,
and sort the contributions into gauge-invariant subsets with
simpler analytic properties than the full amplitude.

\begin{figure}
\begin{center}
\includegraphics[width=4.0in]{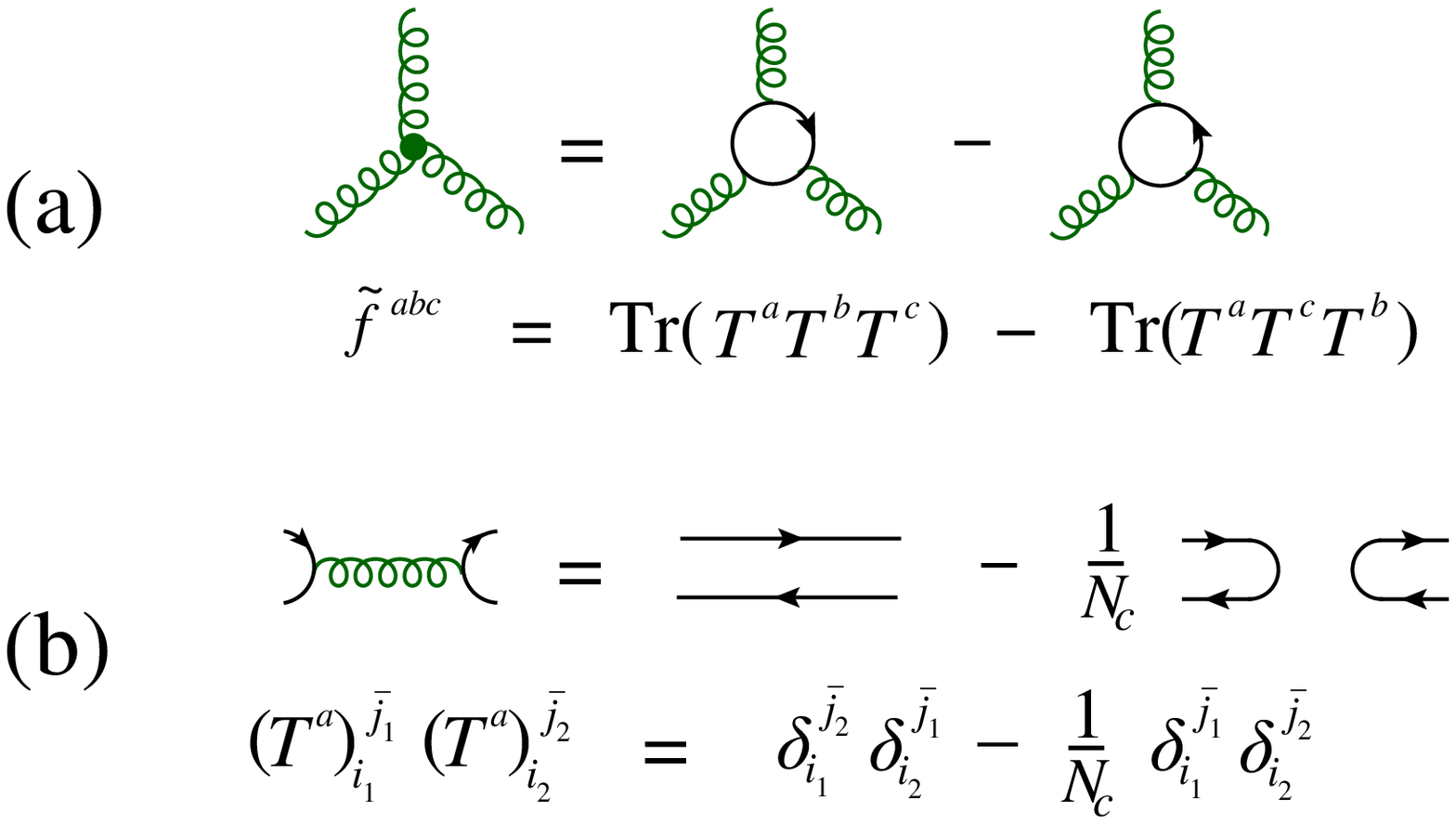}
\end{center}
\caption{Graphical representation of (a) the identity for eliminating
structure constants $f^{abc}$ and (b) the $SU(N_c)$ Fierz identity for
simplifying the resulting traces.}
\label{ColordiagFigure}
\end{figure}

To do this, we first eliminate all the structure constants $f^{abc}$
in favor of the generators $T^a$, using
\be \label{struct}
\tilde{f}^{abc}\ \equiv\ i\sqrt2 f^{abc}
\ =\ \Tr\bigl(T^aT^bT^c\bigr) - \Tr\bigl(T^aT^cT^b\bigr),
\ee
which follows from the definition~(\ref{fabcdef})
of the structure constants.
This identity is represented graphically in \fig{ColordiagFigure}(a),
in which curly lines are in the adjoint representation and lines
with arrows are in the fundamental representation.
After this step, every color factor for a multi-gluon amplitude
is a product of some number of traces.  Many traces share
$T^a$'s with contracted indices, of the form 
$\Tr\bigl(\ldots T^a\ldots\bigr)\,\Tr\bigl(\ldots T^a\ldots\bigr)
\,\ldots\,\Tr\bigl(\ldots)$.
If external quarks are present, then in addition to the traces there
will be some strings of $T^a$'s terminated by fundamental indices,
of the form $(T^{a_1}\ldots T^{a_m})_{i_2}^{~\ib_1}$. 
In order to reduce the number of traces and strings we can
apply the $SU(N_c)$ Fierz identity,
\be \label{colorfierz}
  (T^a)_{i_1}^{~\jb_1} \, (T^a)_{i_2}^{~\jb_2}\ =\ 
  \delta_{i_1}^{~\jb_2} \delta_{i_2}^{~\jb_1}
  - {1\over N_c} \, \delta_{i_1}^{~\jb_1} \delta_{i_2}^{~\jb_2}\,,
\ee
where the sum over $a$ is implicit.  This identity
is illustrated graphically in \fig{ColordiagFigure}(b).

\Eqn{colorfierz} is just the statement that the $SU(N_c)$
generators $T^a$ form the complete set of traceless hermitian 
$N_c\times N_c$ matrices.  The $-1/N_c$ term implements the
tracelessness condition.  (To see this, contract both sides
of \eqn{colorfierz} with $\delta^{~i_1}_{\jb_1}$.) 
It is often convenient to consider also
$U(N_c)\ =\ SU(N_c) \times U(1)$ gauge theory.  The additional $U(1)$ 
generator is proportional to the identity matrix, 
\be \label{photongenerator}
(T^{a_{U(1)}})_{i}^{~\jb}\ =\ {1 \over \sqrt{N_c}}\ \delta_{i}^{~\jb}\ ;
\ee
when this generator is included in the sum over $a$ in \eqn{colorfierz},
the corresponding $U(N_c)$ result is \eqn{colorfierz} without the
$-1/N_c$ term.   
The auxiliary $U(1)$ gauge field is often referred to as a photon.
It is colorless, commuting with $SU(N_c)$, with vanishing
structure constants $f^{a_{U(1)}bc}=0$ for all $b,c$.
Therefore it does not couple directly to gluons, although
quarks carry charge under it.  Real photon amplitudes can be obtained
using this generator, after replacing factors of the strong coupling $g$
with the QED coupling $\sqrt{2}e$.
 
\begin{figure}
\begin{center}
\includegraphics[width=5.0in]{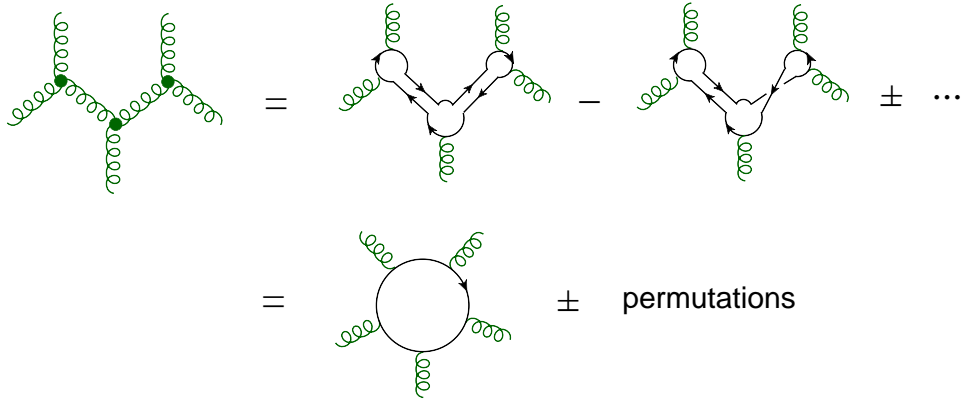}
\end{center}
\caption{Graphical illustration of reducing the color factor for
a five-gluon Feynman diagram to a single color trace.}
\label{FiveptcolorFigure}
\end{figure}

The color algebra can easily be carried out graphically~\cite{DoubleLine},
as illustrated in \fig{FiveptcolorFigure}.
Starting with any given Feynman diagram, one interprets it
as just the color factor for the full diagram, after expanding the 
four-gluon vertices into two three-gluon vertices.
Then one makes the two substitutions, \eqns{struct}{colorfierz}, 
which are represented diagrammatically in \fig{ColordiagFigure}.
In \fig{FiveptcolorFigure} we use these steps to
simplify a sample diagram for five-gluon scattering at tree level.
Inserting the rule \fig{ColordiagFigure}(a) in the three vertices leads
to $2^3 = 8$ terms, of which two are shown in the first line.
The $SU(N_c)$ Fierz identity takes the traces of products of three 
$T^a$'s, and systematically combines them into a single trace,
$\Tr\bigl(T^{a_1}T^{a_2}T^{a_3}T^{a_4}T^{a_5}\bigr)$, plus all possible
permutations, as shown in the second line of the figure.
Notice that the $-1/N_c$ term in \eqn{colorfierz} and
\fig{ColordiagFigure}(b) does not
contribute here, because the photon does not couple to gluons;
that is, $f^{abI} = 0$ when $I$ is the $U(1)$ generator.
(The $-1/N_c$ term only has to be retained when a gluon can couple to
a fermion line at both ends.)

From \fig{FiveptcolorFigure} it is clear that any tree diagram for
$n$-gluon scattering can be reduced to a sum of ``single trace'' terms,
in which the generators $T^{a_i}$ corresponding to the external gluons
have different cyclic orderings.
The {\it color decomposition} of the the $n$-gluon
tree amplitude~\cite{TreeColor} is,
\be \label{treegluecolor}
 {\cal A}^\tree_n ( \{k_i,\lambda_i,a_i\} )
\ =\ g^{n-2} \hskip-1.3mm \sum_{\sigma \in S_n/Z_n} \hskip-1.3mm  
    \Tr( T^{a_{\sigma(1)}}\ldots T^{a_{\sigma(n)}} )\ 
     A_n^\tree(\sigma(1^{\lambda_1}),\ldots,\sigma(n^{\lambda_n})). 
\ee
Here $g$ is the gauge coupling ($g^2/(4\pi)=\alpha_s$), 
$k_i, \lambda_i$ are the gluon momenta and helicities, 
and $A_n^\tree(1^{\lambda_1},\ldots,n^{\lambda_n})$ are the
{\it partial amplitudes}, which contain all the kinematic information.
$S_n$ is the set of all permutations of $n$ objects, while $Z_n$ is
the subset of cyclic permutations, which preserves the trace;
one sums over the set $S_n/Z_n$ in order to sweep out all
cyclically-inequivalent orderings in the trace.
We write the helicity label for each particle, $\lambda_i=\pm$,
as a superscript.

The real work is in calculating the independent partial amplitudes
$A_n^\tree$. However, they are simpler than the full amplitude because
they are  {\it color-ordered}:  they only receive contributions from 
diagrams with a particular cyclic ordering of the gluons.
This feature reduces the number of singularities they can
contain.  Tree amplitudes contain factorization poles, when
a single intermediate state goes on its mass shell in the middle of the
diagram. The momentum of the intermediate state
is the sum of a number of the external momenta.
In the color-ordered partial amplitudes, those momenta must
be cyclically adjacent in order to produce a pole.
For example, the five-point partial amplitudes
$A_5^\tree(1^{\lambda_1},2^{\lambda_2},3^{\lambda_3},
4^{\lambda_4},5^{\lambda_5})$ can only have poles in 
$s_{12}$, $s_{23}$, $s_{34}$, $s_{45}$, and $s_{51}$,
and not in $s_{13}$, $s_{24}$, $s_{35}$, $s_{41}$, or $s_{52}$,
where $s_{ij} \equiv (k_i+k_j)^2$. 
Similarly, at the loop level, only the channels made out of sums of
cyclically adjacent momenta will have unitarity cuts (as well as
factorization poles). The number of cyclically-adjacent momentum
channels grows much more slowly than the total number of channels,
as the number of legs increases.  Later we will use factorization
properties to construct tree amplitudes, so defining partial amplitudes
with a minimal number of factorization channels will simplify the
construction.

Although we have mainly considered the pure-gluon case, color decompositions
can be found for generic QCD amplitudes.  Another simple case is the
set of tree amplitudes $\qb qgg\ldots g$ with two external
quarks, which can be reduced to single strings of $T^a$ matrices,
\be \label{treequarkgluecolor}
 {\cal A}^\tree_n 
\ =\ g^{n-2} \sum_{\sigma \in S_{n-2}} 
   ( T^{a_{\sigma(3)}}\ldots T^{a_{\sigma(n)}} )_{i_2}^{~\jb_1} 
     A_n^\tree(1_\qb^{\lambda_1},2_q^{\lambda_2},
      \sigma(3^{\lambda_3}),\ldots,\sigma(n^{\lambda_n})), 
\ee
where numbers without subscripts refer to gluons. 
Color decompositions for tree amplitudes with more than two external
quarks can be found in ref.~\cite{MPReview}.

The same ideas also work at the loop level~\cite{LoopColor}.
For example, at one loop, the same graphical analysis leads to
a color decomposition for pure-gluon amplitudes which contains
two types of terms:
\begin{itemize}
\item single-trace terms, of the form
$N_c\, \Tr( T^{a_1}\ldots T^{a_n})$ plus permutations,
which contain an extra factor of $N_c$ and dominate for large $N_c$, and
\item double-trace terms, of the form
$\Tr( T^{a_1}\ldots T^{a_m})\Tr(T^{a_{m+1}}\ldots T^{a_n})$ plus permutations,
whose contribution to the color-summed cross section is suppressed
by at least a factor of $1/N_c^2$ with respect to the leading-color
terms.
\end{itemize}
Quark loops lead to contributions of the first type, but with an over all
factor of the number of light quark flavors, $n_f$, replacing the factor
of $N_c$.

After we have computed all of the partial amplitudes,
the parton model requires us to construct the squared
amplitude, averaged over the colors
of the initial-state partons, and summed over the final-state colors.
Using the above color decompositions, and applying Fierz identities again,
this color-summed cross section can be expressed in terms of the partial
amplitudes.  The color factors that appear can be computed graphically.
Take a single trace structure of the type shown in
\fig{FiveptcolorFigure}, and glue the $n$ gluon lines to a second 
trace structure from the conjugate amplitude, which may have a relative
permutation.  Then apply the Fierz identity in
\fig{ColordiagFigure}(b) to remove the gluon lines and
reduce the resulting ``vacuum'' color graph to powers of $N_c$.
(A closed loop for an arrowed line gives a factor of $\Tr(1) = N_c$.)

In this way you can show that the
color-summed cross section for $n$-gluon scattering,
\be
d\sigma^\tree\ \propto\ \sum_{a_i=1}^{N_c^2-1} | \cA^\tree_n(\{k_i,a_i\}) |^2 \,,
\label{colorsum1}
\ee
takes the form,
\be
d\sigma^\tree\ \propto\ N_c^n \bigg\{ \sum_{\sigma\in S_n/Z_n}
\Bigl|  A_n^\tree(\sigma(1),\sigma(2),\ldots,\sigma(n)) \Bigr|^2
+ {\cal O}(1/N_c^2) \biggr\} \,.
\label{colorsum2}
\ee
In other words, the leading-color contributions come from gluing
together two trace structures with no relative permutation, which gives
rise to a planar vacuum color graph.  Any relative permutation leads
to a nonplanar graph, and its evaluation results in at least two fewer
powers of $N_c$.  Of course these subleading-color terms can be worked
out straightforwardly as well.
Another way of stating \eqn{colorsum2} is that, up to $1/N_c^2$-suppressed
terms, the differential cross section can be written as a sum
of positive terms, each of which has a definite color flow.
This description is important for the development of parton showers,
which exploit the pattern of radiating additional soft gluons
from these color-ordered pieces of the cross section.

\section{The spinor helicity formalism}
\label{sec:helicity}

\subsection{Spinor variables}
\label{sec:spinvar}

Now we turn from color to spin.  That is, we ask how to organize
the spin quantum numbers of the external states in order to simplify
the calculation.  The answer is that the helicity basis is a very 
convenient one for most purposes.  In high-energy collider processes, almost
all fermions are ultra-relativistic, behaving as if they were massless.
Massless fermions that interact through gauge interactions have a
conserved helicity, which we can exploit by computing in the helicity
basis.  Although vector particles like photons and gluons do not have
a conserved helicity, it turns out that the most helicity-violating
processes one can imagine are zero at tree level (due to a hidden
supersymmetry that relates boson and fermion amplitudes).  Also,
the nonzero amplitudes are ordered in complexity by how much helicity
violation they have; we will see that the so-called maximally helicity
violating (MHV) amplitudes are the simplest, the next-to-MHV are the next
simplest, and so on.

A related question is, what are the right kinematic variables
for scattering amplitudes?  It is traditional to use the four-momenta,
$k_i^\mu$, and especially their Lorentz-invariant products, 
$s_{ij} = (k_i+k_j)^2$, as the basic kinematic variables.
However, all the particles in the Standard Model --- except the Higgs
boson --- have spin, and for particles with spin, there is a better
choice of variables.  Just as we rewrote the color factors for
$SU(N_c)$ adjoint states ($f^{abc}$) in terms of those associated with the 
smaller fundamental representation of $SU(N_c)$ ($T^a$),
we should now consider trading the
Lorentz vectors $k_i^\mu$ for kinematic quantities that transform under
a smaller representation of the Lorentz group.  

The only available smaller representation of the Lorentz group is the
spinor representation, which for massless vectors can be
two-dimensional (Weyl spinors).  So we trade the four-momentum $k_i^\mu$ 
for a pair of spinors,
\be
k_i^\mu \quad\qquad \Rightarrow \quad\qquad 
u_+(k_i) \equiv |i^+\rangle \equiv \lambda_i^\alpha \,,
\qquad
u_-(k_i) \equiv |i^-\rangle \equiv \lambda_i^{\dot\alpha} \,.
\label{kforlambda}
\ee
Here $u_+(k_i) = \hf(1+\gamma_5)u(k_i)$ is a right-handed spinor
written in four-component Dirac notation, and $\lambda_i^\alpha$ is its
two-component version, $\alpha=1,2$.
Similarly, $u_-(k_i) = \hf(1-\gamma_5)u(k_i)$
is a left-handed spinor in Dirac notation, and $\tilde\lambda_i^\alpha$
is the two-component version, $\dot\alpha=1,2$.
We also give the ``ket'' notation that is often used.
The massless Dirac equation is satisfied by these spinors,
\be
\ksl_i u_\pm(k_i)\ =\ \ksl_i | i^\pm\rangle\ =\ 0.
\label{MasslessDirac}
\ee
There are also negative-energy solutions $v_\pm(k_i)$, but
for $k_i^2=0$ they are not distinct from $u_\mp(k_i)$.
The undotted and dotted spinor indices correspond to two
different spinor representations of the Lorentz group.

We would like to build some Lorentz-invariant quantities out of
the spinors, which we can do using the antisymmetric tensors
$\pol^{\alpha\beta}$ and $\pol^{\dot\alpha\dot\beta}$.  We define the
spinor products,
\bea
\spa{i}.{j} \ &\equiv&\ 
\pol^{\alpha\beta} (\lambda_i)_\alpha (\lambda_j)_\beta 
\ =\ \bar{u}_-(k_i) u_+(k_j), \label{spadef}\\
\spb{i}.{j} \ &\equiv&\
\pol^{\dot\alpha\dot\beta} (\tilde\lambda_i)_{\dot\alpha} (\lambda_j)_{\dot\beta} 
\ =\ \bar{u}_+(k_i) u_-(k_j), \label{spbdef}
\eea
where we give both the two- and four-component versions.

Recall the form of the positive energy projector for $m=0$:
\be
u_+(k_i)\bar{u}_+(k_i)\ =\ |i^+\rangle\,\langle i^+|
\ =\ \hf(1+\gamma_5) \ksl_i \, \hf(1-\gamma_5).
\label{posEproj}
\ee
In two-component notation, this relation becomes,
using the explicit form of the Pauli matrices,
\be
(\lambda_i)_\alpha (\tilde\lambda_i)_{\dot\alpha} 
\ =\ k_i^\mu (\sigma_\mu)_{\alpha\dot\alpha}
\ =\ (\ksl_i)_{\alpha\dot\alpha}
\ =\ \left( 
\begin{matrix} k_i^t + k_i^z & k_i^x - i k_i^y \\
               k_i^x + i k_i^y & k_i^t - k_i^z \\
\end{matrix}
\right).
\label{posEproj2}
\ee
Note that the determinant of this $2\times2$ matrix vanishes,
$\det(\ksl_i) = k_i^2 = 0$, which is consistent with its
factorization into a column vector $(\lambda_i)_\alpha$
times a row vector $(\tilde\lambda_i)_{\dot\alpha}$.

Also note that if the momentum vector $k_i^\mu$ is real, then complex
conjugation is equivalent to transposing the matrix $\ksl_i$,
which via \eqn{posEproj2}
corresponds to exchanging the left- and right-handed spinors,
$(\tilde\lambda_i)_{\dot\alpha} \lr (\lambda_i)_\alpha$.
In other words, for real momenta, a chirality flip of
all spinors (which could be induced by a parity transformation)
is the same as complex conjugating the spinor products,
\be
\spb{i}.{j}\ =\ \spa{i}.{j}^\ast \,.
\label{conjugation}
\ee

If we contract \eqn{posEproj2} with $(\sigma^\nu)^{\dot\alpha\alpha}$,
we find that we can reconstruct the four-momenta $k_i^\mu$
from the spinors,
\be
\langle i^+ | \gamma^\mu | i^+ \rangle\ \equiv\
(\tilde\lambda_i)_{\dot\alpha} (\sigma^\mu)^{\dot\alpha\alpha} (\lambda_i)_\alpha
\ =\ 2 k_i^\mu \,.
\label{recon4mom}
\ee
Using the Fierz identity for Pauli matrices,
\be
(\sigma^\mu)_{\alpha\dot\alpha} (\sigma_\mu)^{\dot\beta\beta}
\ =\ 2 \delta_\alpha^\beta \delta_{\dot\alpha}^{\dot\beta} \,,
\label{PauliFierz}
\ee
we can similarly reconstruct the momentum invariants from the spinor
products,
\be
2 k_i\cdot k_j \, = \, \frac{1}{2}  
(\tilde\lambda_i)_{\dot\alpha} (\sigma^\mu)^{\dot\alpha\alpha} (\lambda_i)_\alpha
(\tilde\lambda_j)_{\dot\beta} (\sigma_\mu)^{\dot\beta\beta} (\lambda_j)_\beta
\, = \, (\lambda_i)_\alpha (\lambda_j)^\alpha 
  (\tilde\lambda_j)_{\dot\alpha} (\tilde\lambda_i)^{\dot\alpha} \,,
\label{squaring1}
\ee
or
\be
s_{ij}\ =\ \spa{i}.{j} \spb{j}.{i} \,.
\label{squaring2}
\ee

For real momenta, we can combine \eqns{conjugation}{squaring2} to
see that the spinor products are complex square roots
of the momentum-invariants,
\be
\spa{i}.{j} = \sqrt{s_{ij}} e^{i\phi_{ij}}\,, \qquad
\spb{i}.{j} = \sqrt{s_{ij}} e^{-i\phi_{ij}}\,,
\label{spinorsquareroots}
\ee
where $\phi_{ij}$ is some phase.  We will see later that this
complex square-root property allows the spinor products to capture
perfectly the singularities of amplitudes as two massless momenta
become parallel (collinear).  This fact is one way of understanding
why helicity amplitudes can be so compact when written in terms
of spinor products.

We collect here some useful spinor product identities:
\bea
&&\hbox{anti-symmetry}:\quad \spa{i}.{j} = -\spa{j}.{i}, 
\quad \spb{i}.{j} = -\spb{j}.{i}, 
\quad \spa{i}.{i} = \spb{i}.{i} = 0, \label{antisym}\\
&&\hbox{squaring}:\quad \spa{i}.{j} \spb{j}.{i} = s_{ij},
\label{squaring}\\
&&\hbox{momentum\ conservation}:\quad \sum_{j=1}^n \spa{i}.{j} \spb{j}.{k} = 0
\label{momcons}\\
&&\hbox{Schouten}:\quad \spa{i}.{j} \spa{k}.{l}
- \spa{i}.{k} \spa{j}.{l} = \spa{i}.{l} \spa{k}.{j}.
\label{schouten}
\eea
Note also that the massless Dirac equation in two-component notation
follows from the antisymmetry of the spinor products:
\be
(\ksl_i)_{\dot\alpha\alpha} (\lambda_i)^\alpha 
\ =\ (\tilde\lambda_i)_{\dot\alpha} \spa{i}.{i}\ =\ 0.
\label{MasslessDirac2}
\ee
Finally, for numerical evaluation it is useful to have explicit
representations of the spinors,
\be
(\lambda_i)_\alpha = \left(
\begin{matrix}
\sqrt{k_i^t + k_i^z} \\
\frac{k_i^x + i k_i^y}{\sqrt{k_i^t + k_i^z}} \\
\end{matrix}
\right) \,,
\qquad
(\tilde\lambda_i)_{\dot\alpha} = \left(
\begin{matrix}
\sqrt{k_i^t + k_i^z} \\
\frac{k_i^x - i k_i^y}{\sqrt{k_i^t + k_i^z}} \\
\end{matrix}
\right) \,,
\label{explicitspinors}
\ee
which satisfy \eqns{posEproj2}{conjugation}.

We would like to have the same formalism describe amplitudes that
are related by crossing symmetry, {\it i.e.}, by moving various
particles between the initial and final states.  In order to keep
everything on a crossing-symmetric footing, we define the momenta
as if they were all outgoing, so that initial-state momenta are
assigned the negative of their physical momenta.  Then momentum conservation
for an $n$-point process takes the crossing symmetric form,
\be
\sum_{i=1}^n k_i^\mu = 0.
\label{momncons}
\ee
We also label the helicity as if the particle were outgoing.
For outgoing particles this label is the physical helicity, but
for incoming particles it is the opposite.  Because of this,
whenever we look at a physical pole of an amplitude, and assign
helicities to an intermediate on-shell particle, the helicity
assignment will always be opposite for amplitudes appearing on 
two sides of a factorization pole.
The same consideration will apply to particles crossing a cut,
at the loop level.


\subsection{A simple four-point example}
\label{sec:4pthel}

\begin{figure}
\begin{center}
\includegraphics[width=3.0in]{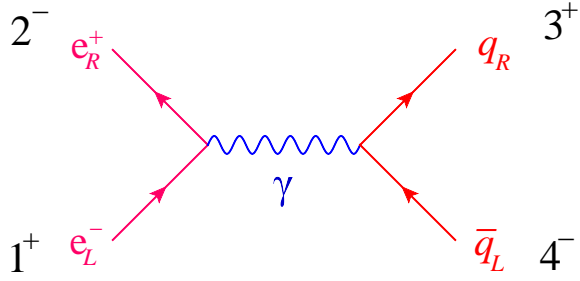}
\end{center}
\caption{The one Feynman diagram for $e^-e^+ \to q\qb$. 
Particles are labeled with $L$ and $R$ subscripts for left- and
right-handed particles.  We also give in black the numerical, all-outgoing
labeling convention.}
\label{eeqqFigure}
\end{figure}

Let's illustrate spinor-helicity methods with the simplest
scattering amplitude of all, the one for electron-positron
annihilation into a massless fermion pair, say a pair of quarks,
for which the single Feynman diagram is shown in \fig{eeqqFigure}.
This amplitude is related by crossing symmetry
to the amplitude for electron-quark scattering at the core of deep
inelastic scattering, and by time reversal symmetry to the
annihilation of a quark and anti-quark into a pair of leptons,
{\it i.e.}~the Drell-Yan reaction.

We take all the external states to be helicity eigenstates,
choosing first to consider,
\be
e_L^-(-k_1) e_R^+(-k_2) \to q_R(k_3) \qb_L(k_4)\,.
\label{eeqq}
\ee
Note that we have assigned momenta $-k_1$ and $-k_2$ to the 
incoming states, so that momentum conservation takes the
crossing-symmetric form,
\be
k_1+k_2+k_3+k_4 = 0.
\label{mom4cons}
\ee
In the all-outgoing helicity labeling convention, the incoming left-handed
electron is labeled as if it were an outgoing right-handed positron
(positive-helicity $\bar{e}$),
and similarly for the incoming right-handed positron
(labeled as a negative-helicity $e$).
We label the amplitude with numerals $i$ standing for the momenta $k_i$,
subscripts to identify the type of particle (if it is not a gluon),
and superscripts to indicate the helicity.  Thus the amplitude for
reaction~(\ref{eeqq}) is denoted by
\be
\cA_4^\tree(1_{\bar{e}}^+,2_e^-,3_q^+,4_\qb^-) \equiv {\cal A}_4 \,.
\label{notation4}
\ee

As discussed above, we first strip off the color factors, as well
as any other coupling factors.  In this case the color factor is
a trivial Kronecker $\delta$ that equates the quark colors.
We define the color-stripped amplitude $A_4$ by
\be
\cA_4 = (\sqrt{2}e)^2 Q_e Q_q \delta_{i_3}^{\ib_4} \, A_4 \,,
\label{strip4}
\ee
where $e$ is the electromagnetic coupling, obeying
$e^2/(4\pi) = \alpha_{\rm QED}$, and $Q_e$ and $Q_q$ are the electron
and quark charges.  The factor of $(\sqrt{2}e)^2$ arises because
it is convenient to normalize the color-stripped
amplitudes so that there are no $\sqrt{2}$ factors for QCD.
In this normalization, the substitution $g\to\sqrt{2}e$ is required
in the prefactor for each QED coupling. A corresponding $(1/\sqrt2)^2$
goes into the Feynman rule for $A_4$.

The usual Feynman rules for the diagram in~\fig{eeqqFigure} give
\bea
A_4 &=& \frac{i}{2s_{12}} \overline{v_{-}}(k_2)\gamma^\mu u_{-}(k_1)
           \overline{u_{+}}(k_3) \gamma_\mu v_+(k_4) \nn\\
    &=& \frac{i}{2s_{12}} (\sigma^\mu)_{\alpha\dot\alpha} (\lambda_2)^\alpha
       (\tilde{\lambda}_1)^{\dot\alpha} (\sigma_\mu)^{\dot\beta\beta}
       (\tilde{\lambda}_3)_{\dot\beta} (\lambda_4)_\beta \,,
\label{evaluateA4A}
\eea
where we have switched to two-component notation in the second line.
Now we apply the Fierz identity for Pauli matrices,
\eqn{PauliFierz}, obtaining
\be
A_4\ =\ \frac{i}{s_{12}} (\lambda_2)^\alpha (\tilde{\lambda}_1)^{\dot\alpha} 
               (\lambda_4)_\alpha (\tilde{\lambda}_3)_{\dot\alpha}
   \ =\ i \, \frac{\spa2.4 \spb1.3}{s_{12}} \,,
\label{evaluateA4B}
\ee
after using the definitions~(\ref{spadef}) and (\ref{spbdef})
of the spinor products $\spa{i}.{j}$ and $\spb{i}.{j}$.

According to \eqns{squaring}{conjugation},
the spinor products are square-roots of the
momentum invariants, up to a phase.  Because $s_{24}=s_{13}$ for massless
four-point kinematics, we can rewrite~\eqn{evaluateA4B} as
\be
A_4\ =\ i \, \frac{\spa2.4 \spb1.3}{s_{12}}
  \ =\ e^{i\phi} \frac{s_{13}}{s_{12}}\ =\ -\frac{e^{i\phi}}{2} (1-\cos\theta) \,,
\label{rewriteA4}
\ee
where $\phi$ is some phase angle,
and $\theta$ is the center-of-mass scattering angle.
From this formula, we can check the helicity suppression of the amplitude
in the forward scattering limit, $A_4\to0$  as $\theta\to0$.
The amplitude vanishes in this limit because of angular-momentum
conservation:  the initial angular momentum
along the $e_L^{-}$ direction is $(-\hf)-\hf = -1$, while
the final angular momentum is $\hf-(-\hf) = +1$.
At $\theta=\pi$, the spins line up and there is no suppression.

The result~(\ref{evaluateA4B}) for $A_4$ is in a mixed representation,
containing both the ``holomorphic'' (right-handed) spinor product $\spa2.4$
and the ``anti-holomorphic'' (left-handed) spinor product $\spb1.3$.
However, we can multiply top and bottom by $\spa{1}.{3}$, and use 
the squaring relation~(\ref{squaring}), $s_{13}=s_{24}$ and momentum
conservation~(\ref{momcons}) to rewrite it as,
\be
A_4 = i\, \frac{\spa2.4\spb1.3}{s_{12}} 
    = i\, \frac{\spa2.4\spb1.3\spa1.3}{\spa1.2\spb2.1 \spa1.3} 
    = - i\, \frac{\spa2.4\spb2.4\spa2.4}{\spa1.2\spb2.4 \spa4.3} 
    = i\, \frac{{\spa2.4}^2}{\spa1.2\spa3.4} \,.
\label{A4toholo}
\ee
The latter form only involves the spinors $\spa{i}.{j}$.
On the other hand, the same identities also allow us to write it in an
anti-holomorphic form.  In summary, we have
\be
A_4^\tree(1_{\bar{e}}^+,2_e^-,3_q^+,4_\qb^-) 
    = i\, \frac{{\spa2.4}^2}{\spa1.2\spa3.4} 
    = i\, \frac{{\spb1.3}^2}{\spb1.2\spb3.4} \,.
\label{A4pmpm}
\ee
It turns out that $A_4^\tree(1_{\bar{e}}^+,2_e^-,3_q^+,4_\qb^-)$ is
the first in an infinite series of ``maximally helicity violating'' 
(MHV) amplitudes, containing these four fermions along with $(n-4)$
additional positive-helicity gluons or photons.  All of these MHV
amplitudes, containing exactly two negative-helicity particles,
are holomorphic.  (We will compute one of them in a little while.)
But $A_4^\tree(1_{\bar{e}}^+,2_e^-,3_q^+,4_\qb^-)$ is
also the first in an infinite series of $\overline{\rm MHV}$ amplitudes,
containing these four fermions along with $(n-4)$
additional {\it negative}-helicity gluons or photons.
All the $\overline{\rm MHV}$ amplitudes are anti-holomorphic;
in fact, they are the parity conjugates of the MHV amplitudes.
As a four-point amplitude, \eqn{A4pmpm} has a dual life,
belonging to both the MHV and the $\overline{\rm MHV}$ series.
The same phenomenon occurs for other classes of amplitudes, including
the $n$-gluon MHV amplitudes (the Parke-Taylor amplitudes~\cite{ParkeTaylor})
and their $\overline{\rm MHV}$ conjugate amplitudes, which we will
encounter shortly.

So far we have only computed one helicity configuration for 
$e^+e^-\to q\qb$.  There are 16 configurations in all.  However,
the helicity of massless fermions is conserved when they interact
with gauge fields, or in the all-outgoing labeling, the positron's helicity
must be the opposite of the electron's, and the antiquark's helicity
must be the opposite of the quark's.  So there are only $2\times2 = 4$
nonvanishing helicity configurations.  They are all related by parity (P)
and by charge conjugation (C) acting on one of the fermion lines.
For example, C acting on the electron line exchanges labels 1 and 2,
which can also be interpreted as flipping the helicities of particles
1 and 2, taking us from \eqn{A4pmpm} to
\be
A_4^\tree(1_{\bar{e}}^-,2_e^+,3_q^+,4_\qb^-) 
    = - i\, \frac{{\spa1.4}^2}{\spa1.2\spa3.4} \,.
\label{A4mppm}
\ee
Parity flips all helicities and conjugates all spinors,
$\spa{i}.{j} \to \spb{i}.{j}$, taking us from \eqn{A4pmpm} to
\be
A_4^\tree(1_{\bar{e}}^-,2_e^+,3_q^-,4_\qb^+) 
    = i\, \frac{{\spb2.4}^2}{\spb1.2\spb3.4} \,.
\label{A4mpmp}
\ee
Combining the two operations leads to
\be
A_4^\tree(1_{\bar{e}}^+,2_e^-,3_q^-,4_\qb^+) 
    = - i\, \frac{{\spb1.4}^2}{\spb1.2\spb3.4} \,.
\label{A4pmmp}
\ee
Of course eqs.~(\ref{A4mppm}), (\ref{A4mpmp}) and (\ref{A4pmmp})
can all be rewritten in the conjugate variables as well.

The scattering probability, or differential cross section,
is proportional to the square of the amplitude.
Squaring a single helicity amplitude would give the cross section for fully
polarized incoming and outgoing particles.
In QCD applications, we rarely have access to the spin states of the partons.
Hadron beams are usually unpolarized, so the incoming quarks and gluons
are as well.  The outgoing quarks and gluons
shower and fragment to produce jets of hadrons, wiping out almost all
traces of final-state parton helicities.  In other words, we need
to construct the unpolarized cross section, by summing over all
possible helicity configurations. (The different helicity configurations
do not interfere with each other.)  For our $e^+e^- \to q\qb$ example,
we need to sum over the four nonvanishing helicity configurations,
after squaring the tree-level helicity amplitudes.
The result, omitting the overall coupling and flux factors, is
\bea
\frac{d\sigma}{d\cos\theta}\ \propto\ \sum_{\rm hel.} |A_4|^2
\ &=&\ 2 \, \biggl\{ \biggl| \frac{{\spa2.4}^2}{\spa1.2\spa3.4} \biggr|^2
            + \biggl| \frac{{\spa1.4}^2}{\spa1.2\spa3.4} \biggr|^2 \biggr\}
\nn\\
\ &=&\ 2 \, \frac{ s_{24}^2 + s_{14}^2 }{ s_{12}^2 } \nn\\
\ &=&\ \frac{1}{2} \Bigl[ (1-\cos\theta)^2 + (1+\cos\theta)^2 \Bigr] \nn\\
\ &=&\ 1 + \cos^2\theta.
\label{unpolcross}
\eea
We used the fact that the amplitudes related by parity are equal
up to a phase, in order to only exhibit two of the four nonzero helicity
configurations explicitly.

For a simple process like $e^+e^- \to q\qb$, helicity amplitudes
are overkill.  It would be much faster to use the textbook method of
computing the unpolarized differential cross section directly,
by squaring the amplitude for generic external spinors and
using Casimir's trick of inserting the positive energy projector
for the product of two spinors, summed over spin states.
The problem with this method is that the computational effort scales
very poorly when there a large number of external legs $n$.  The number of
Feynman diagrams grows like $n!$, so the number of separate interferences
between diagrams in the squared amplitude goes like $(n!)^2$.  That is
why all modern methods for high-multiplicity scattering processes
compute amplitudes, not cross sections, for some basis of external
polarization states.  For massless particles, this is usually the helicity
basis.  After computing numerical values for the helicity amplitudes
at a given phase-space point, the cross section is constructed from
the helicity sum.

\subsection{Helicity formalism for massless vectors}
\label{sec:helvec}

Next we consider external massless vector particles,
{\it i.e.}~gluons or photons.
Spinor-helicity techniques began in the early 1980s with the
recognition~\cite{SpinorHelicity} that polarization vectors for
massless vector particles with definite helicity could be constructed
from a pair of massless spinors, as follows:
\bea
(\pol_i^+)_\mu = \pol_\mu^+(k_i,q) 
&=& \frac{\sandmm{q}.{\gamma_\mu}.{i}}
{\sqrt{2} \spa{q}.{i}} \,,
\qquad
(\pol_i^-)_\mu = \pol_\mu^-(k_i,q) 
= - \frac{\sandpp{q}.{\gamma_\mu}.{i}}
{\sqrt{2} \spb{q}.{i}} \,, 
\label{helpol}\\
(\esl_i^+)_{\alpha\dot\alpha} = \, \esl_{\alpha\dot\alpha}^+(k_i,q)
&=& \sqrt{2} \frac{ \lambda_q^\alpha \tilde{\lambda}_i^{\dot\alpha} }
          {\spa{q}.{i}} \,,
\qquad\ \ \
(\esl_i^-)_{\alpha\dot\alpha} = \, \esl_{\alpha\dot\alpha}^-(k_i,q)
= - \sqrt{2} \frac{ \tilde{\lambda}_q^{\dot\alpha} \lambda_i^\alpha } 
          {\spb{q}.{i}} \,,
\label{helpolsl}
\eea
where we have also given the $2\times2$ matrix version, from contracting
with a $\sigma$ matrix and using the Fierz identity~(\ref{PauliFierz}).
Here $k_i^\mu$ is the gluon momentum and $q^\mu$ is an additional massless
vector called the reference momentum, whose associated two-component
left- and right-handed spinors are $\tilde{\lambda}_q^{\dot\alpha}$
and $\lambda_q^\alpha$.  Using the massless Dirac equation,
\be
\ksl_i |i^\pm\rangle\ =\ 0\ =\ \qsl | q^\pm \rangle \,,
\label{MasslessDirac3}
\ee
we see that the polarization vectors~(\ref{helpol}) obey the required
transversality with respect the gluon momentum,
\be
\pol_i^\pm \cdot k_i = 0\,.
\label{requiredtransv}
\ee
As a bonus, it also is transverse with respect to $q$:\ 
$\pol_i^\pm\cdot q=0$.

The second form~(\ref{helpolsl}) for the polarization vector shows that
$\esl_i^+$ produces a state with helicity $+1$, because it contains two
complex conjugate spinors with momentum $k_i$ in the numerator and denominator.
These two spinors pick up opposite spin-$1/2$ phases under an
azimuthal rotation about the $k_i$ axis,
\be
\tilde\lambda_i^{\dot\alpha} \to e^{i\phi/2} \tilde\lambda_i^{\dot\alpha} \,,
\qquad \lambda_i^{\alpha} \to e^{-i\phi/2} \lambda_i^{\alpha} \,,
\label{aziphase}
\ee
so the ratio transforms like helicity $+1$,
\be
\esl_i^+ \propto \frac{\tilde\lambda_i^{\dot\alpha}}{ \lambda_i^{\alpha}}
\ \to\ e^{i\phi} \esl_i^+ \,.
\label{aziphase2}
\ee

There is a freedom to choose different reference vectors $q_i$ for each
of the external states $i$.  This freedom is the residual on-shell gauge
invariance, that amplitudes should be unchanged when the polarization
vector is shifted by an amount proportional to the momentum.
A judicious choice of the reference vectors can greatly simplify
a Feynman diagram computation by causing many diagrams to vanish.
However, we won't be doing many Feynman diagram computations, just
the one in the next subsection, of a five-point amplitude. In this case,
there are only two diagrams, one of which we will make
vanish through a choice of $q$.

\subsection{A five-point amplitude}

In this subsection we compute one of the next simplest helicity amplitudes,
the one for producing a gluon along with the quark-antiquark pair
in $e^+e^-$ annihilation.  This amplitude contributes to three-jet
production in $e^+e^-$ annihilation, and to the next-to-leading order
corrections to deep inelastic scattering and to Drell-Yan production,
in the crossed channels.

\begin{figure}
\begin{center}
\includegraphics[width=5.5in]{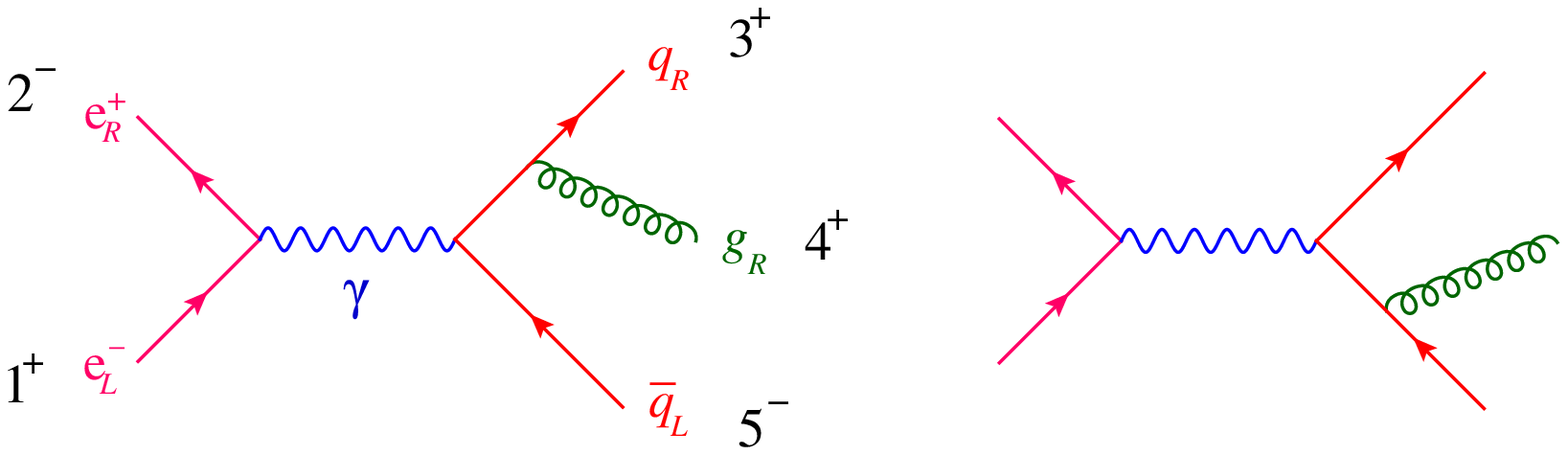}
\end{center}
\caption{The two Feynman diagrams for $e^-e^+ \to qg\qb$.}
\label{eeqgqFigure}
\end{figure}

We compute the amplitude for the helicity configuration
\be
e_L^-(-k_1) e_R^+(-k_2) \to q_R(k_3) g_R(k_4) \qb_L(k_5)\,,
\label{eeqgq}
\ee
namely
\be
\cA_5^\tree(1_{\bar{e}}^+,2_e^-,3_q^+,4^+,5_\qb^-) \equiv {\cal A}_5 \,.
\label{notation5}
\ee
Again we strip off the color and charge factors, defining
\be
\cA_5 = (\sqrt{2}e)^2 g \, Q_e Q_q (T^{a_4})_{i_3}^{\ib_5} \, A_5 \,,
\label{strip5}
\ee
where $A_5$ is constructed from the two Feynman diagrams in
\fig{eeqgqFigure}.
%

Recall that in the evaluation of the four-point
amplitude~(\ref{evaluateA4B}), after applying the Fierz identity
related to the photon propagator,
the two external fermions with the same (outgoing) helicity had
their spinors contracted together, generating factors of $\spa2.4$ and
$\spb1.3$.  In the two diagrams in \fig{eeqgqFigure}, the same
thing happens for the quark or anti-quark that does not have a gluon
emitted off it, generating a factor of $\spa2.5$ in the first diagram
and $\spb1.3$ in the second one.  On the other spinor string, we have
to insert a factor of the off-shell fermion propagator and the gluon
polarization vector, giving
\be
A_5 = - i\, \frac{\spa2.5}{s_{12}} \frac{\sandpm{1}.{(\ksl_3+\ksl_4)\esl_4^+}.{3}}
  {\sqrt{2} s_{34}}
 + i\, \frac{\spb1.3}{s_{12}} \frac{\sandmp{2}.{(\ksl_4+\ksl_5)\esl_4^+}.{5}}
  {\sqrt{2} s_{45}} \,.
\label{eval5ptA}
\ee
Inserting the formula~(\ref{helpolsl}) for the gluon polarization vector,
we obtain
\be
A_5 = - i\, \frac{\spa2.5}{s_{12}} 
\frac{\sandpp{1}.{(\ksl_3+\ksl_4)}.{q} \spb4.3}
  {s_{34} \spa{q}.{4}}
 + i\, \frac{\spb1.3}{s_{12}} \frac{\sandmm{2}.{(\ksl_4+\ksl_5)}.{4} \spa{q}.5}
  {s_{45} \spa{q}.{4}} \,.
\label{eval5ptB}
\ee
Now we choose the reference momentum $q=k_5$ in order to make the second
graph vanish,
\be
A_5 = - i\, \frac{\spa2.5}{s_{12}} \frac{\sandpp{1}.{(\ksl_3+\ksl_4)}.{5}\spb4.3}
  {s_{34} \spa5.4}
    = - i\, \frac{\spa2.5\spb1.2\spa2.5\spb4.3}
             {\spa1.2\spb2.1\spa3.4\spb4.3\spa4.5}
    = i\, \frac{{\spa2.5}^2}{\spa1.2\spa3.4\spa4.5} \,,
\label{eval5ptC}
\ee
where we used momentum conservation~(\ref{momcons}) and a couple
of other spinor-product identities to simplify the answer to its
final holomorphic form,
\be
A_5(1_{\bar{e}}^+,2_e^-,3_q^+,4^+,5_\qb^-)
 = i\, \frac{{\spa2.5}^2}{\spa1.2\spa3.4\spa4.5} \,.
\label{A5}
\ee
(As an exercise in spinor-product identities, verify \eqn{A5}
for other choices of $q$.)

Next we will study the behavior of $A_5$ in various
kinematic limits, which will give us insight into the generic singular
behavior of QCD amplitudes.

\section{Soft and collinear factorization}
\label{sec:factorization}

In this section, we use the five-point amplitude~(\ref{A5})
to verify some universal limiting behavior of QCD amplitudes.  In the next
section, we will use this universal behavior to derive recursion
relations for general tree amplitudes.

\subsection{Soft gluon limit}

\begin{figure}
\begin{center}
\includegraphics[width=4.5in]{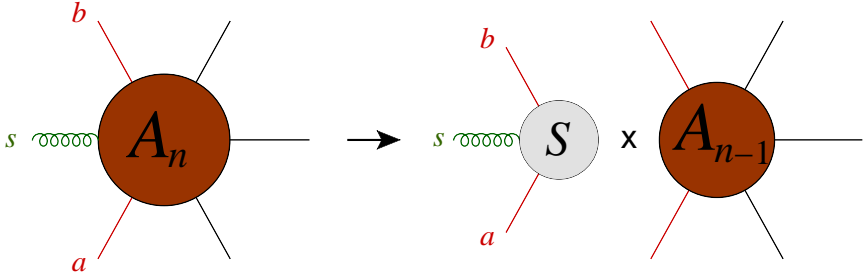}
\end{center}
\caption{Factorization of a QCD amplitude when a soft gluon $s$ is
emitted between the hard partons $a$ and $b$.}
\label{SoftfactFigure}
\end{figure}

First consider the limit that the gluon momentum $k_4$ in \eqn{A5}
becomes soft, {\it i.e.}~scales uniformly to zero, $k_4\to0$.
In this limit, we can {\it factorize} the amplitude into a divergent piece
that depends on the energy and angle of the emitted gluon, and a second piece
which is the amplitude omitting that gluon:
\bea
A_5(1_{\bar{e}}^+,2_e^-,3_q^+,4^+,5_\qb^-) 
= i\, \frac{{\spa2.5}^2}{\spa1.2\spa3.4\spa4.5}
 &=& \frac{\spa3.5}{\spa3.4\spa4.5}
  \times i\, \frac{{\spa2.5}^2}{\spa1.2\spa3.5} \nn\\
&\to& {\cal S}(3,4^+,5) \times A_4(1_{\bar{e}}^+,2_e^-,3_q^+,5_\qb^-) \,.
\label{A5soft}
\eea
The {\it soft factor} (or eikonal factor) is given more
generally by,
\be
{\cal S}(a,s^+,b) = \frac{\spa{a}.{b}}{\spa{a}.{s}\spa{s}.{b}} \,,
\qquad
{\cal S}(a,s^-,b) = -\frac{\spb{a}.{b}}{\spb{a}.{s}\spb{s}.{b}} \,,
\label{softfactor}
\ee
where $s$ labels the soft gluon, and $a$ and $b$ label the two
hard partons that are adjacent to it in the color ordering.

Although we have only inspected the soft limit of one amplitude,
the more general result is,
\be
A_n^\tree(1,2,\ldots,a,s^\pm,b,\ldots,n) \inlimit^{k_s\to0}\
{\cal S}(a,s^\pm,b) \times 
A_{n-1}^\tree(1,2,\ldots,a,b,\ldots,n)\,.
\label{softlimitgen}
\ee
This factorization is depicted in \fig{SoftfactFigure}.\footnote{
Actually, the case we inspected in \eqn{A5soft} was somewhat
special in that we didn't need to use the fact that $k_s\to0$
in order to put the five-point amplitude into the limiting form
of \eqn{softlimitgen}; normally one would have to do so.}
The $(n-1)$-point amplitude on the right-hand side is that obtained
by just deleting the soft-gluon $s$ in the $n$-point amplitude.
The soft factor is universal:  it does not depend on whether $a$ and $b$
are quarks or gluons; it does not care about their helicity;
and it does not even depend on the magnitude of their momenta, just
their angular direction (as one can see by rescaling the spinor $\lambda_a$
in \eqn{softfactor}).  The spin independence arises because soft emission
is long-wavelength, and intrinsically classical.  Because of this,
we can pretend that the external partons $a$ and $b$ are scalars, and
compute the soft factor simply from two Feynman diagrams, from
emission off legs $a$ and $b$.  We can use the
scalar QED vertex in the numerator, while the (singular) soft limit of the
adjacent internal propagator generates the denominator:
\be
{\cal S}(a,s^+,b) = - \frac{\sqrt{2} \pol_s^+(q)\cdot k_a}{2k_a\cdot k_s}
             + \frac{\sqrt{2} \pol_s^+(q)\cdot k_b}{2k_b\cdot k_s}
 = \frac{\spa{a}.{q}}{\spa{s}.{q}\spa{a}.{s}}
 - \frac{\spa{b}.{q}}{\spa{s}.{q}\spa{b}.{s}}
 = \frac{\spa{a}.{b}}{\spa{a}.{s}\spa{s}.{b}} \,,
\label{derivesoft}
\ee
using the Schouten identity~(\ref{schouten}) in the last step.


\subsection{Collinear limits}
\label{sec:collimits}

Next consider the limit of the $e^+e^-\to qg\bar{q}$ amplitude~(\ref{A5})
as the quark momentum $k_3\equiv k_a$ and the gluon momentum 
$k_4\equiv k_b$ become
parallel, or collinear.  This limit is singular because the
intermediate momentum $k_P \equiv k_a + k_b$ is going on shell
in the collinear limit:
\be
k_P^2 = 2k_a\cdot k_b \inlimit^{a\parallel b} 0.
\label{coll_ab}
\ee
We also need to specify the relative longitudinal momentum fractions carried
by partons $a$ and $b$, 
\be
k_a \approx z k_P \,, \qquad k_b \approx (1-z) k_P \,,
\label{zfracmom}
\ee
where $0<z<1$.
This relation implies, thanks to \eqn{explicitspinors}, that the spinors
obey similar relations with square roots:
\bea
&&\lambda_a \approx \sqrt{z} \, \lambda_P \,, 
\qquad \lambda_b \approx \sqrt{1-z} \, \lambda_P \,,
\label{zfracspa}\\
&&\tilde\lambda_a \approx \sqrt{z} \, \tilde\lambda_P \,, 
\qquad \tilde\lambda_b \approx \sqrt{1-z} \, \tilde\lambda_P \,,
\label{zfracspb}
\eea
%

\begin{figure}
\begin{center}
\includegraphics[width=4.5in]{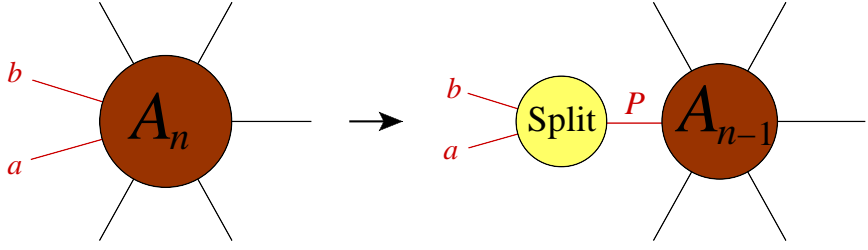}
\end{center}
\caption{Factorization of a QCD amplitude when two color-adjacent
partons $a$ and $b$ become collinear.}
\label{CollfactFigure}
\end{figure}

Inserting \eqn{zfracspa} into \eqn{A5}, we find that
\bea
A_5(1_{\bar{e}}^+,2_e^-,3_q^+,4^+,5_\qb^-) 
= i\, \frac{{\spa2.5}^2}{\spa1.2\spa3.4\spa4.5}
 &\approx& \frac{1}{\sqrt{1-z}\spa3.4}
  \times i\, \frac{{\spa2.5}^2}{\spa1.2\spa{P}.5} \nn\\
&\to& \Split_-(3_q^+,4_g^+;z) \times A_4(1_{\bar{e}}^+,2_e^-,P_q^+,5_\qb^-) \,.
\nn\\ \label{A5coll34}
\eea
Here we have introduced the {\it splitting amplitude}
$\Split_{-\lambda_P}(a^{\lambda_a},b^{\lambda_b};z)$, which governs
the general collinear factorization of tree amplitudes
depicted in \fig{CollfactFigure},
\be
 A_{n}^\tree(\ldots,a^{\lambda_a},b^{\lambda_b},\ldots)
\inlimit^{a \parallel b}\
\sum_{\lambda_P=\pm} 
\Split_{-\lambda_P}(a^{\lambda_a},b^{\lambda_b};z)\,
         A_{n-1}^\tree(\ldots,P^{\lambda_P},\ldots) \,.
\label{collgen}
\ee
In contrast to the soft factor, the splitting amplitude depends
on whether $a$ and $b$ are quarks or gluons, and on their helicity.
It also includes a sum over the helicity $\lambda_P$ of the intermediate
parton $P$.  (Note that the labeling of $\lambda_P$ is reversed between
the $(n-1)$-point tree amplitude and the splitting amplitude, because we
apply the all-outgoing helicity convention to the splitting amplitude
as well.)  The $(n-1)$-point tree amplitude on the right-hand side
of \eqn{collgen} is found by merging the two partons, according to the
possible splittings in QCD:  $g\to gg$, $g\to q\bar{q}$, $\bar{q} \to \bar{q}g$
and (in this case) $q\to qg$.  For the splitting amplitude 
$\Split_-(a_q^+,b_g^+;z)$ entering \eqn{A5coll34}, quark helicity conservation
implies that only one of the two intermediate helicities survives.
For intermediate gluons, both signs of $\lambda_P$ can appear in general.
As in the case of the soft limit, the four-point amplitude $A_4$ is found
by relabeling \eqn{A4pmpm}.

One can also extract from \eqn{A5} the splitting amplitude for the case
that the (anti)quark and gluon have the opposite helicity, by taking
the collinear limit $4\parallel5$.  The two results can be summarized
as:
\bea
\Split_-(q^+,g^+) &=& \frac{1}{\sqrt{1-z}\spa{q}.{g}} \,, \qquad
\Split_-(q^+,g^-) = -\frac{z}{\sqrt{1-z}\spb{q}.{g}} \,, \label{qgcoll}\\
\Split_-(g^+,\qb^+) &=& \frac{1}{\sqrt{z}\spa{g}.{\qb}} \,, \qquad
\Split_-(g^-,\qb^+) = -\frac{1-z}{\sqrt{z}\spb{g}.{\qb}} \,.
\label{gqbcoll}
\eea
where the other cases (including some not shown, with opposite quark
helicity) are related by parity or charge conjugation.

Collinear singularities in the initial state give rise to the DGLAP
evolution equations for parton distributions.  In fact, the splitting
amplitudes are essentially the square root of the (polarized)
Altarelli-Parisi splitting probabilities which are the kernels
of the DGLAP equations.  That is, the $z$ dependence
of the splitting amplitudes, after squaring and summing over the helicities
$\lambda_a$, $\lambda_b$ and $\lambda_P$, reproduces the
splitting probabilities.  For example, one can reconstruct the correct
$z$-dependence of the $q\to qg$ splitting probabilities $P_{qq}(z)$
using \eqn{qgcoll}, squaring and summing over the gluon helicity:
\be
P_{qq}(z)\ \propto\ \left(\frac{1}{\sqrt{1-z}}\right)^2
              + \left(\frac{z}{\sqrt{1-z}}\right)^2
\ =\ \frac{1+z^2}{1-z} \,,
\label{Pqq}
\ee
while $P_{gq}(z)$ is given by exchanging $z \lr 1-z$.
\Eqn{Pqq} omits the $\delta(1-z)$ term from virtual gluon emission, but
its coefficient can be inferred from quark number conservation.

\subsection{The Parke-Taylor amplitudes}
\label{sec:PT}

In the all-outgoing helicity convention, one can show that the
pure-gluon amplitudes for which all the gluon helicities are the same,
or at most one is different from the rest, vanish for any $n\geq4$:
\be
A_n^\tree(1^\pm,2^+,\ldots,n^+) = 0.
\label{treevanish}
\ee
(Cyclic symmetry allows us to move a single negative-helicity gluon to leg 1.)
This result can be proven directly by noticing that the tree amplitude
contains $n$ different polarization vectors, contracted together 
with at most $n-2$ momenta (because there are at most $n-2$ cubic vertices
in any Feynman graph, each of which is linear in the momentum).
Therefore every term in every tree amplitude contains at least one
polarization vector contraction of the form $\pol_i\cdot\pol_j$.
Inspecting the form of the polarization vectors in \eqn{helpol},
we see that like-helicity contractions, $\pol_i^+(q_i)\cdot\pol_j^+(q_j)$,
vanish if $q_i=q_j$, while opposite helicity contractions,
$\pol_i^-(q_i)\cdot\pol_j^+(q_j)$, vanish if $q_i=k_j$ or $q_j=k_i$.
To show that $A_n^\tree(1^+,2^+,\ldots,n^+)$ vanishes, we can
just choose all reference momenta to be the same, $q_i=q$.
To show that $A_n^\tree(1^-,2^+,\ldots,n^+)$ vanishes, we can choose
$q_i=k_1$ for $i>1$ and $q_1 = k_2$, for example.
It is also possible to prove \eqn{treevanish} using the fact that
tree-level $n$-gluon amplitudes are the same in QCD as in a supersymmetric
theory~\cite{NewSWI}, and so they obey Ward identities for
supersymmetric scattering amplitudes~\cite{OldSWI}.

The remarkable simplicity of gauge-theory scattering amplitudes
is encapsulated by the Parke-Taylor~\cite{ParkeTaylor}
amplitudes for the MHV $n$-gluon amplitudes, in which exactly two
gluons, $j$ and $l$, have opposite helicity from the rest:
\be 
\label{mhvall}
A_{jl}^{\rm MHV}\ \equiv\ 
A_n^\tree(1^+,\ldots,j^-,\ldots,l^-,\ldots,n^+)\ =\ 
  i\, { {\spa{j}.{l}}^4 \over \spa1.2\cdots\spa{n}.{1} }\ .
\ee

One of the reasons these amplitudes are so simple is that they
have no multi-particle poles --- no factors of 
$1/(k_m+k_{m+1}+\cdots+k_p)^2 \equiv 1/P^2$
for $p>m+1$.  Why is that?  A multi-particle pole would correspond
to factorizing the scattering process into two subprocesses, each
with at least four gluons,
\be
A_n^\tree(\ldots)\ \inlimit^{P^2\to0}\ A_{n-k+1}(\ldots,P^{\lambda_P},\ldots)
 \frac{i}{P^2} A_{k+1}(\ldots,(-P)^{-\lambda_P},\ldots) \,, \quad
3\leq k\leq n-3,
\label{multiparticle}
\ee
In the MHV case, there are two negative-helicity gluons among the 
arguments ``$\ldots$'' of the two tree amplitudes on the
right-hand side of \eqn{multiparticle}, plus one more for either
$P$ or $(-P)$ (but not both).  That's three negative-helicity gluons
to be distributed among two tree amplitudes. However, \eqn{treevanish}
says that both trees need at least two negative helicities to be nonvanishing,
for a minimum of four required.  Hence the multiparticle poles
must all vanish, due to insufficiently many negative helicities.
As we'll see in \sect{sec:loops}, similar arguments control the
structure of loop amplitudes as well.

We have found that the MHV amplitudes have no multi-particle factorization
poles, consistent with \eqn{mhvall}.  Their principal singularities are
the soft and collinear limits.  It's easy to check that the soft
limit~(\ref{softlimitgen}) is satisfied by the MHV amplitudes
in \eqn{mhvall}.  It's also simple to verify that the collinear
behavior~(\ref{collgen}) is obeyed, and to extract the $g\to gg$ splitting
amplitudes,
\bea
\Split_-(a^+,b^+) &=& \frac{1}{\sqrt{z(1-z)} \spa{a}.{b}} \,,
\quad
\Split_+(a^-,b^+) = \frac{z^2}{\sqrt{z(1-z)} \spa{a}.{b}} \,,
\nn\\
\Split_+(a^+,b^-) &=& \frac{z^2}{\sqrt{z(1-z)} \spa{a}.{b}} \,,
\quad \Split_+(a^+,b^+) = 0 \,,
\label{splitggg}
\eea
plus their parity conjugates.  The last relation in \eqn{splitggg} must
hold for consistency, because otherwise the collinear limit of an MHV
amplitude (which has no multi-particle poles)
could generate a next-to-MHV amplitude with three negative helicities
(which generically does have such poles).
It's a useful exercise to reconstruct the unpolarized $g\to gg$ splitting
probabilities $P_{gg}(z)$ from \eqn{splitggg} by squaring and summing over
all helicity configurations.

A closely related series of MHV amplitudes to the pure-glue ones
are those with a single external 
$q\bar{q}$ pair and $(n-2)$ gluons.  In this case helicity conservation along
the fermion line forces either the quark or antiquark to have negative
helicity. Using charge conjugation, we can pick it to be the antiquark.
Referring to the color decomposition~(\ref{treequarkgluecolor}),
the partial amplitudes for which all gluons have the same helicity vanish
identically,
\be
A_n^\tree(1_\qb^-,2_q^+,3^+,4^+,\ldots,n^+) = 0,
\label{treeqgvanish}
\ee
while the MHV ones with exactly one negative-helicity gluon (leg $i$)
take the simple form,
\be 
\label{mhvqgall}
A_n^\tree(1_\qb^-,2_q^+,3^+,\ldots,i^-,\ldots,n^+)\ =\ 
  i\, { {\spa{1}.{i}}^3 \spa{2}.{i} \over \spa1.2\cdots\spa{n}.{1} }\,.
\ee
It's easy to see that the absence of multi-particle poles in \eqn{mhvall},
whether for intermediate gluons or quarks,
again follows from the vanishing relations~(\ref{treevanish}) and
(\ref{treeqgvanish}), and simple counting of negative helicities.
However, the relation between the pure-glue
MHV amplitudes~$A_{1i}^{\tree,{\rm MHV}}$ in \eqn{mhvall} and the
quark-glue ones~(\ref{mhvqgall}) is much closer than that,
as they differ only by a factor of $\spa{2}.{i}/\spa{1}.{i}$.
These relations follow from supersymmetry Ward
identities~\cite{OldSWI,NewSWI,MPReview,LDTASI95}.

\subsection{Spinor magic}

\begin{figure}
\begin{center}
\includegraphics[width=4in]{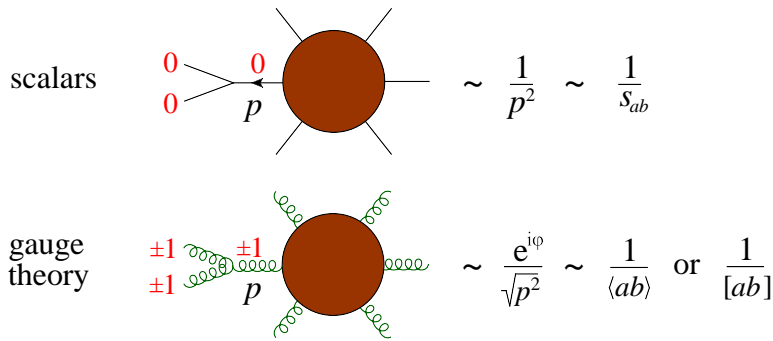}
\end{center}
\caption{In gauge theory, an angular-momentum mismatch lessens
the singular behavior from $1/p^2$ to $1/\sqrt{p^2}$, 
and introduces an azimuthally-dependent phase, both of which are captured
by the spinor products.}
\label{SpinorMagicFigure}
\end{figure}

All of the splitting amplitudes contain denominator factors of either
$\spa{a}.{b}$ or its parity conjugate $\spb{a}.{b}$.
From \eqn{spinorsquareroots}, we see that the collinear singularity
is proportional to the square root of the momentum invariant that is
vanishing, times a phase.  This phase varies as the two collinear partons
are rotated in the azimuthal direction about their common axis.
Both the square root and the phase behavior follow from angular momentum
conservation in the collinear limit.  \Fig{SpinorMagicFigure}
illustrates the difference between scalar $\phi^3$ theory and gauge theory.
In scalar $\phi^3$ theory, no spin angular momentum is carried by either the
external scalars or the intermediate one.  Thus there is no violation of
angular-momentum conservation along the collinear axis.
Related to this, the three-vertex shown carries no momentum dependence,
and the collinear pole is determined solely by the scalar propagator to
be $\sim 1/s_{ab}$ in the limit that legs $a$ and $b$ become parallel.

In contrast, in every collinear limit in massless gauge theory, angular
momentum conservation is violated by at least one unit.  In the pure-glue
case shown in \fig{SpinorMagicFigure}, the intermediate
physical gluon must be transverse and have helicity $\pm1$, but this value
is never equal to the sum of the two external helicities: $\pm1\pm1=\pm2$ or 0.
The helicity mismatch forces the presence of orbital angular momentum,
which comes from the momentum dependence in the gauge-theory three-vertex.
It suppresses the amplitude in the collinear limit, from
$1/s_{ab}$ to $1/\sqrt{s_{ab}}$, similarly to the vanishing of
$A_4$ in \eqn{rewriteA4} in the limit $\theta\to0$. 
The helicity mismatch also generates the azimuthally-dependent phase.
The sign of the mismatch, by $\pm1$ unit, is correlated with whether the
splitting amplitude contains $1/\spa{a}.{b}$ or $1/\spb{a}.{b}$, since
these spinor products acquire opposite phases under an azimuthal rotation.

In summary, the spinor products are the perfect variables for capturing
the collinear behavior of massless gauge theory amplitudes, simply due
to angular-momentum considerations.  Because collinear singularities
dictate many of the denominator factors that should appear in the analytic
representations of amplitudes, we can now understand more physically
why the spinor product representation can lead to such compact analytic
results.

\subsection{Complex momenta, spinor products and three-point kinematics}
\label{sec:kin3pt}

There is another reason the spinor products are essential for 
modern amplitude methods, and that is to make sense out of
massless three-point scattering.  If we use only momentum invariants,
then the three-point kinematics, defined by
\be
k_1^\mu + k_2^\mu + k_3^\mu = 0, \qquad k_1^2 = k_2^2 = k_3^2 = 0,
\label{kin3ptgen}
\ee
is pathological.  For example, $s_{12} = (k_1+k_2)^2 = k_3^2 = 0$, 
and similarly every momentum invariant $s_{ij}$ vanishes.
If the momenta are real, then \eqn{spinorsquareroots} implies
that all the spinor products vanish as well, $\spa{i}.{j}=\spb{i}.{j}=0$.
It is easy to see that for real momenta the only solutions to 
\eqn{kin3ptgen} consist of strictly parallel four-vectors,
which is another way of seeing why all dot products and spinor
products must vanish.

However, if the momenta are complex, there is a loophole:
The conjugation
relation~(\ref{conjugation}), $\spb{i}.{j} = \spa{i}.{j}^\ast$,
does not hold, although the relation~(\ref{squaring2}),
$s_{ij} = \spa{i}.{j} \spb{j}.{i}$, is still true.
Therefore we can have some of the spinor products be nonzero,
even though all the momentum invariants vanish, $s_{ij}=0$.
There are two chirally conjugate solutions:
\begin{enumerate}
\item
$\tilde{\lambda}_1 \propto \tilde{\lambda}_2 \propto\tilde{\lambda}_3\quad
 \Rightarrow\quad$all $\spb{i}.{j}=0$ while all $\spa{i}.{j}\neq0$.
\item
$\lambda_1 \propto \lambda_2 \propto \lambda_3\quad
 \Rightarrow\quad$all $\spa{i}.{j}=0$ while all $\spb{i}.{j}\neq0$.
\end{enumerate}
The proportionality of the two-component spinors causes the corresponding
spinor products to vanish.  There are no continuous variables associated
with the three-point process, so one should think of the kinematical
region as consisting of just two points, which are related to each
other by parity.

For the first choice of kinematics, MHV three-point amplitudes such as
\be
A_3^\tree(1^-,2^-,3^+) = i \frac{{\spa1.2}^4}{\spa1.2\spa2.3\spa3.1}
\label{MHV3}
\ee
make sense and are nonvanishing.
$\overline{\rm MHV}$ three-point amplitudes such as
\be
A_3^\tree(1^+,2^+,3^-) = -i \frac{{\spb1.2}^4}{\spb1.2\spb2.3\spb3.1}
\label{MHVbar3}
\ee
are nonvanishing for the second type of kinematics.  When
the MHV three-point amplitudes are nonvanishing, the $\overline{\rm MHV}$
ones vanish, and vice versa.

It's important to note that the splitting amplitudes defined in
\sect{sec:collimits} correspond to approximate three-point kinematics
with real momenta, whereas the three-point amplitudes~(\ref{MHV3})
and (\ref{MHVbar3}) correspond to exact three-point kinematics with
complex momenta.  They are similar notions, but not exactly the same thing.

\vfill\eject


\section{The BCFW recursion relation for tree amplitudes}
\label{sec:trees}

\subsection{General formula}

The idea behind the derivation of the BCFW recursion
relation~\cite{BCFW} is that tree-level amplitudes are plastic,
or continuously deformable, analytic functions of the scattering
momenta.  Therefore, it should be possible to reconstruct amplitudes 
for generic scattering kinematics from their behavior in singular limiting
kinematics.  In these singular regions, amplitudes split, or factorize,
into two causally disconnected amplitudes with fewer legs, connected
by a single intermediate state, which can propagate an arbitrary
distance because it is on its mass shell. 

Multi-leg amplitudes depend on many variables, and multi-variable
complex analysis can be tricky.  However, BCFW considered a
family of on-shell tree amplitudes, $A_n(z)$, depending on a single complex
parameter $z$ which shifts some of the momenta. 
(We drop the ``tree'' superscript here for convenience.)
This family explores enough of the singular kinematical
configurations to allow recursion relations to be derived for
the original amplitude at $z=0$, $A_n = A_n(0)$.
There have since been many generalizations of this approach, leading to
different types of recursion relations.  The BCFW momentum shift
only affects two of the momenta, say legs $n$ and $1$.
The shift can be defined using the spinor variables as,
\bea
\tilde\lambda_n &\to& \hat{\tilde\lambda}_n
    = \tilde\lambda_n - z\tilde\lambda_1 \,,
\qquad \lambda_n \to \lambda_n \,, \nn\\
\lambda_1 &\to& \hat{\lambda}_1 = \lambda_1 + z\lambda_n \,,
\qquad \tilde\lambda_1 \to \tilde\lambda_1 \,,
\label{BCFWshift}
\eea
where hatted variables indicate variables after the shift.
This particular shift is called the $[n,1\rangle$ shift, because it
only affects the spinor products involving the left-handed spinor 
$\tilde\lambda_n$ and the right-handed spinor $\lambda_1$.

The shift~(\ref{BCFWshift}) can also be expressed in terms
of momentum variables,
\bea
\hat{\ksl}_1(z)\ &=&\ 
(\lambda_1 + z \lambda_n) \, \tilde\lambda_1 
\ =\ \lambda_1\tilde\lambda_1 + z \lambda_n \tilde\lambda_1 \,,
\nn\\
\hat{\ksl}_n(z)\ &=&\ 
\lambda_n \, (\tilde\lambda_n - z \tilde\lambda_1)
\ =\ \lambda_n\tilde\lambda_n - z \lambda_n \tilde\lambda_1 \,,
\label{BCFWmomshift}
\eea
which makes clear that momentum conservation holds for any value of $z$,
because
\be
\hat{k}_1^\mu(z) + \hat{k}_n^\mu(z) = k_1^\mu + k_n^\mu \,.
\label{BCFWmomcons}
\ee
Also, since both $\hat{\ksl}_1(z)$ and $\hat{\ksl}_n(z)$ in
\eqn{BCFWmomshift} can be factorized as $2\times2$ matrices
into row vectors times column vectors, their determinants vanish.
Then, according to the discussion
around \eqn{posEproj2}, they remain on shell,
\be
\hat{k}_1^2(z) =  \hat{k}_n^2(z) = 0.
\label{BCFWonshell}
\ee
We can give a physical picture of the direction of the momentum shift
by first writing $\hat{k}_1^\mu(z) = k_1^\mu + z v^\mu$,
$\hat{k}_n^\mu(z) = k_1^\mu - z v^\mu$.  Requiring \eqn{BCFWonshell}
for all $z$ implies that $v\cdot k_1 = v\cdot k_n = v^2 = 0$.
If we go to a Lorentz frame in which the spatial components of $k_1$
and $k_n$ are both along the $z$ direction, then we see that $v^\mu$ must
be a null vector in the space-like transverse $(x,y)$ plane.
This is only possible if $v^\mu$ is a complex vector.
It's easy to see that $v^\mu = \tfrac{1}{2}\sandpp{1}.{\gamma^\mu}.{n}$
satisfies the required orthogonality relations.

\begin{figure}
\begin{center}
\includegraphics[width=5.5in]{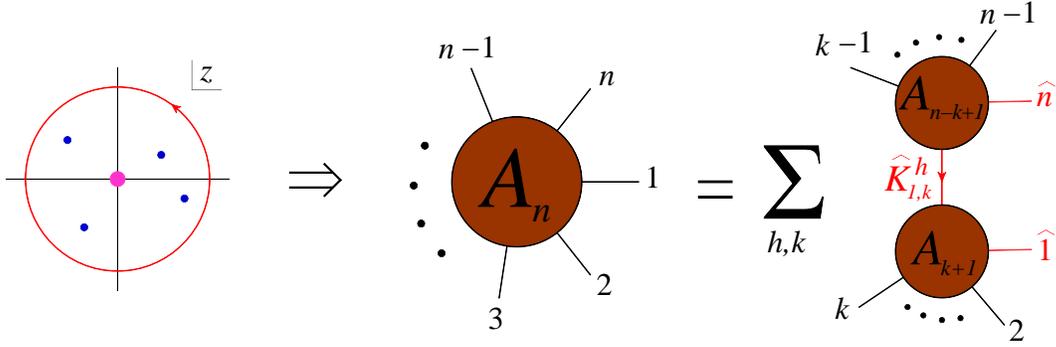}
\end{center}
\caption{Illustration of how Cauchy's theorem leads to the BCFW
recursion relation.  The magenta dot represents the residue at the origin;
the blue dots the residues at $z_k$.  In the recursion relation,
the red lines carry complex, shifted momenta.}
\label{BCFWFigure}
\end{figure}

The function $A_n(z)$ depends meromorphically on $z$.  If it behaves well
enough at infinity, then we can use Cauchy's theorem to relate its behavior
at $z=0$ (the original amplitude) to its residues at finite values
of $z$ (the factorization singularities).  If $A_n(z)\to0$ as $z\to\infty$,
then we have,
\be
0 = \frac{1}{2\pi i} \oint_C dz \frac{A_n(z)}{z}
  = A_n(0) + \sum_k {\rm Res} \biggl[ \frac{A_n(z)}{z} \biggr] \bigg|_{z=z_k} 
\,,
\label{Cauchy1}
\ee
where $C$ is the circle at infinity, and $z_k$ are the locations
of the factorization singularities in the $z$ plane.  (See \fig{BCFWFigure}.)
These poles occur when the amplitude factorizes into a subprocess
with momenta $(\hat{k}_1,k_2,\ldots,k_k,-\hat{K}_{1,k})$,
where $\hat{K}_{1,k}(z_k) = \hat{k}_1(z_k) + k_2 + \cdots + k_k$ must be
on shell.  This information lets us write a simple equation for $z_k$,
\be
0 = \hat{K}^2_{1,k}(z_k) = (\hat{k}_1(z_k) + k_2 + \cdots + k_k)^2
 = (z_k \lambda_n\tilde\lambda_1 + K_{1,k})^2
 = z_k \sandmm{n}.{\Ksl_{1,k}}.{1} + K^2_{1,k} \,,
\label{zkequation}
\ee
where $K_{1,k} = k_1 + k_2 + \cdots + k_k$.
The solution to \eqn{zkequation} is
\be
z_k = - \frac{K^2_{1,k}}{\sandmm{n}.{\Ksl_{1,k}}.{1}} \,.
\label{zk}
\ee
%

We also have to compute the residue of $A(z)/z$ at $z=z_k$.
To do that we use \eqn{multiparticle}, which also holds for three-point
factorizations in complex kinematics.  The singular factor
in the denominator that produces the residue is
\be
zP^2(z) = z \hat{K}^2_{1,k}(z) 
\approx z_k \sandmm{n}.{\Ksl_{1,k}}.{1} (z-z_k)
\approx  - K_{1,k}^2 \, (z-z_k).
\label{getresidue}
\ee
Thus after taking the residue it contributes a factor of the
corresponding scalar propagator,
$i/K_{1,k}^2$, evaluated for the original unshifted kinematics where
it is nonsingular.

Solving \eqn{Cauchy1} for $A_n(0)$ then gives the final BCFW
formula~\cite{BCFW},
\be
A_n(1,2,\ldots,n) = \sum_{h=\pm} \sum_{k=2}^{n-2}
 A_{k+1}(\hat{1},2,\ldots,k,-\hat{K}_{1,k}^{-h})
 \frac{i}{K_{1,k}^2}
 A_{n-k+1}(\hat{K}_{1,k}^h,k+1,k+2,\ldots,n-1,\hat{n}),
\label{BCFWfinal}
\ee
where the hat in the $k^{\rm th}$ term indicates that the shifted momentum
is to be evaluated for $z=z_k$, and $h=\pm$ labels the sign of the helicity
of the intermediate state carrying (complex) momentum $\hat{K}_{1,k}$.
The sum is over the $n-3$ ordered partitions of the $n$ momenta into
two sets, with at least a three-point amplitude on the left ($k\geq2$)
and also on the right ($k\leq n-2$).  The recursion relation is
depicted in \fig{BCFWFigure}.

\begin{figure}
\begin{center}
\includegraphics[width=2in]{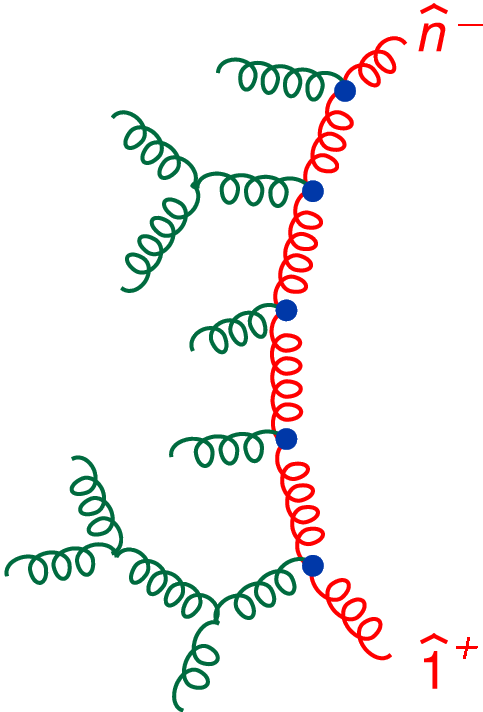}
\end{center}
\caption{Large $z$ dependence of a generic Feynman diagram, for
the $[n^-,1^+\rangle$ momentum shift.  Only the red gluons carry
the large momentum.}
\label{BCFWlargezFigure}
\end{figure}

In order to finish the proof of \eqn{BCFWfinal}, we need to show
that $A_n(z)$ vanishes as $z\to\infty$.  We will do so for the case
that leg $n$ has negative helicity and leg $1$ has positive helicity,
the so-called $[-,+\rangle$ case.  This case can be demonstrated using
Feynman diagrams~\cite{BCFW}.  The cases $[+,+\rangle$ and $[-,-\rangle$
also vanish at infinity, but the proof is slightly more involved.
The case $[+,-\rangle$ diverges at infinity, so it should not be used
as the basis for a recursion relation.  Consider the large $z$ behavior
of the generic Feynman diagram shown in \fig{BCFWlargezFigure}.
Only the red gluons carry the large momentum proportional to $z v^\mu$.
The red propagators contribute factors of the form
\be
\frac{1}{\hat{K}_{1,k}^2(z)}
= \frac{1}{K_{1,k}^2 + z \sandmm{n}.{\Ksl_{1,k}}.{1}}
\sim \frac{1}{z} \,,\qquad \hbox{as }z\to\infty.
\label{largezprop}
\ee
Yang-Mills vertices are (at worst) linear in the momentum,
so they contribute a factor of $z$ per vertex.
There is one more vertex than propagator, so the amplitude
scales like $z^{+1}$ before we take into account the external polarization
vectors.  For the $[-,+\rangle$ case, they scale like,
\be
\esl_n^-(q) \propto \frac{\lambda_n \tilde\lambda_q}{\spb{n}.{q}}
\propto \frac{1}{z} \,,
\qquad
\esl_1^+(q) \propto \frac{\tilde\lambda_1 \lambda_q}{\spa{1}.{q}}
\propto \frac{1}{z} \,.
\label{largezpol}
\ee
The two factors of $1/z$, combined with the factor of $z$ from the internal
part of the diagram, mean that every Feynman diagram falls off like $1/z$,
so $A_n(\infty)=0$ for the $[-,+\rangle$ shift.

It is easy to see that flipping either helicity in \eqn{largezpol}
results in a polarization vector that scales like $z$ instead of $1/z$,
invalidating the argument based on Feynman diagrams.  However, it is
possible to show~\cite{AKlargez}
using the background field method that the
$[+,+\rangle$ and $[-,-\rangle$ cases are actually just as well behaved
as the $[-,+\rangle$ case, also falling off like $1/z$.
In contrast, the $[+,-\rangle$ case does diverge like $z^3$,
as suggested by the above diagrammatic argument.


\subsection{Application to MHV}
\label{sec:BCFWMHVapp}

Next we apply the BCFW recursion relation to prove the form of the
Parke-Taylor amplitudes~(\ref{mhvall}), inductively in the number of
legs $n$. For convenience, we will use cyclicity
to put one of the two negative helicities in the $n^{\rm th}$ position,
\be
\label{mhvalln}
A_{jn}^{\rm MHV}\ \equiv\ 
A_n^\tree(1^+,2^+,\ldots,j^-,\ldots,(n-1)^+,n^-)\ =\ 
  i\, \frac{{\spa{j}.{n}}^4}{\spa1.2\cdots\spa{n}.{1}} \,.
\ee
First we note that the middle terms in the sum over $k$ in \eqn{BCFWfinal},
with $3\leq k \leq n-3$ all vanish.  That's because they correspond
to the multi-particle pole factorizations considered in \eqn{multiparticle},
with at least a four-point amplitude on each side of the factorization
pole, and vanish according to the discussion below \eqn{multiparticle},
by counting negative helicities.

The case $k=n-2$ also vanishes.  If $j=n-1$, then it vanishes because 
$A_{k+1}$ can
have at most one negative helicity.  If $j<n-1$, then we must have $h=+$
so that $A_{k+1}$ is non-vanishing, and then the three-point
amplitude $A_{n-k+1}$ is of type $(+,+,-)$.  This amplitude, given
in \eqn{MHVbar3}, can be nonvanishing when the three right-handed spinors
$\lambda_i$ ($i=K,n-1,n$)
are proportional (the second choice of three-point kinematics).
However, we have shifted the left-handed spinor $\tilde\lambda_n$,
not the right-handed one, and it is easy to check that the three-point
configuration we arrived at is the one for which three left-handed spinors
$\tilde\lambda_i$ are proportional.  For this choice $A_{n-k+1}$ vanishes.

The only nonvanishing contribution is from $k=2$.  We assume $j>2$ for
simplicity. Since we have shifted
$\lambda_1$, the three right-handed spinors $\lambda_i$ ($i=K,1,2$) must
be proportional, which allows the following three-point amplitude
to be non-vanishing:
\be
A_3(\hat{1}^+,2^+,-\hat{K}^-) 
= - i \frac{{\spb{\hat{1}}.{2}}^4}
{[\hat{1}\,2] [2\,(-\hat{K})] [(-\hat{K})\,\hat{1}]}
= + i \frac{{\spb1.2}^3}{[2\,\hat{K}] [\hat{K}\,1]} \,,
\label{BCFWMHV3}
\ee
where $\hat{K} = \hat{K}_{1,2}$.  We removed the hats on 1 in the second
step, since $\tilde\lambda_1$ is not shifted.  There are also two factors
of $i$ from reversing the sign of $\hat{K}$ in the spinor products.

The other amplitude appearing in the $k=2$ term in \eqn{BCFWfinal} is
evaluated using induction on $n$ and \eqn{mhvalln}:
\bea
A_{n-1}(\hat{K}^+,3^+,\ldots,j^-,\ldots,n^-)
&=& i \, \frac{{\spa{j}.{\hat{n}}}^4}{\langle\hat{K}\,3\rangle\spa3.4
   \cdots \spa{n-1,}.{\hat{n}} \langle\hat{n}\,\hat{K}\rangle} \nn\\
&=& i \, \frac{{\spa{j}.{n}}^4}{\langle\hat{K}\,3\rangle\spa3.4
   \cdots \spa{n-1,}.{n} \langle n\,\hat{K}\rangle} \,,
\label{BCFWMHVmany}
\eea
where we can again remove the hats on $n$ because $\lambda_n$ is
unshifted.

Combining the three factors in the $k=2$ term in the BCFW
formula~(\eqn{BCFWfinal}) gives
\be
\label{fullkeq2A}
A_{jn}^{\rm MHV} = - i \frac{{\spa{j}.{n}}^4}{\langle\hat{K}\,3\rangle\spa3.4
   \cdots \spa{n-1,}.{n} \langle n\,\hat{K}\rangle} 
\frac{1}{s_{12}}
 \frac{{\spb1.2}^3}{[2\,\hat{K}] [\hat{K}\,1]} \,.
\ee
One can combine the $\hat{K}$-containing factors into
$\langle n\,\hat{K}\rangle [\hat{K}\,2]$ and
$\langle3\,\hat{K}\rangle [\hat{K}\,1]$.
At this point, we would normally need the value of $z_k$ to proceed.
From \eqn{zk}, it is
\be
z_2 = - \frac{s_{12}}{\sandmm{n}.{(1+2)}.{1}}
 = -\frac{\spa1.2\spb2.1}{\spa{n}.2\spb2.1} = -\frac{\spa1.2}{\spa{n}.2} \,.
\label{zk_keq2}
\ee
However, the evaluation of the $\hat{K}$-containing strings in this
case, where 
\be
\hat{\Ksl} = \hat{\Ksl}_{1,2}(z_2) = \ksl_1 + \ksl_2 
+ z_2 \lambda_n \tilde\lambda_1 \,,
\label{MHVKsl}
\ee
does not actually require the value of $z_2$:
\bea
\langle n\,\hat{K}\rangle [\hat{K}\,2]
&=& \sandmm{n}.{(1+2)}.{2} + z_2 \spa{n}.{n} \spb{1}.{2}
 = \spa{n}.1 \spb1.2\,,
\nn\\
\langle3\,\hat{K}\rangle [\hat{K}\,1]
&=& \sandmm{3}.{(1+2)}.{1} + z_2 \spa3.{n} \spb1.1 = \spa3.2\spb2.1 \,.
\label{MHVidents}
\eea
Inserting these results into \eqn{fullkeq2A} gives
\bea
A_{jn}^{\rm MHV} &=& - i \frac{{\spa{j}.{n}}^4 {\spb1.2}^3}
{(\spa1.2\spb2.1)(\spb1.2\spa2.3)(\spa{n}.1\spb1.2)
 \spa3.4 \cdots \spa{n-1,}.{n}} \nn\\
&=& i \frac{{\spa{j}.{n}}^4}
 {\spa1.2\spa2.3 \cdots \spa{n-1,}.{n} \spa{n}.1} \,,
\label{fullkeq2B}
\eea
completing the induction and proving the Parke-Taylor formula.

\subsection{An NMHV application}
\label{sec:BCFWNMHVapp}

Now we know all the MHV pure-gluon tree amplitudes with exactly
two negative helicities, and by parity, all the $\overline{\rm MHV}$ 
amplitudes with exactly two positive helicities.  The first gluonic
amplitude which is not zero or one of these is encountered for six gluons,
with three negative and three positive helicities, the next-to-MHV
case. In fact, there are three inequivalent cases (up to cyclic permutations
and reflection symmetries):
\be
A_6(1^+,2^+,3^+,4^-,5^-,6^-),\quad
A_6(1^+,2^+,3^-,4^+,5^-,6^-),\quad
A_6(1^+,2^-,3^+,4^-,5^+,6^-).
\label{NMHV6}
\ee
One can use a simple group theory relation known as the $U(1)$
decoupling identity to rewrite the third configuration in terms
of the first two~\cite{MPReview,LDTASI95}.

Here we will give a final illustration of the BCFW recursion relation
by computing the first of the amplitudes in \eqn{NMHV6}.  (The other two
are almost as simple to compute.)
We again use the $[n^-,1^+\rangle$ shift, for $n=6$.
The $k=3$ term vanishes in this case because 
$A_{k+1} = A_4(\hat{1}^+,2^+,3^+,-\hat{K}_{1,3}^{-h}) = 0$.
The $k=2$ and $k=4$ terms are related by the following parity
symmetry:
\be
1\rangle \lr 6], \quad 2\rangle \lr 5], \quad 3\rangle \lr 4], \quad
4\rangle \lr 3], \quad 5\rangle \lr 2], \quad 6\rangle \lr 1].
\label{pppmmmparity}
\ee
For the $k=2$ term, using $z_2$ from \eqn{zk_keq2}, we have the
kinematical identities (where again $\hat{K}=\hat{K}_{1,2}$),
\bea
\hat{\Ksl} &=& \ksl_1 + \ksl_2 - \frac{\spa1.2}{\spa6.2} |6\rangle [1| \,,
\label{pppmmmid1}\\
|\hat{1}] &=& |1] \,,
\label{pppmmmid2}\\
|\hat{6}] &=& |6] + \frac{\spa1.2}{\spa6.2} |1] \,.
\label{pppmmmid3}
\eea
The $k=2$ BCFW diagram is
\bea
T_2 &\equiv& A_3(\hat{1}^+,2^+,-\hat{K}_{1,2}^-)
 \frac{i}{s_{12}}
 A_5(\hat{K}_{1,2}^+,3^+,4^-,5^-,\hat{6}^-) \nn\\
&=& \frac{i}{s_{12}}
 \frac{{\spb{\hat{1}}.{2}}^3}{[2\,\hat{K}] [\hat{K}\,1]}
 \frac{[\hat{K}\,3]^3}
  {\spb3.4\spb4.5\spb{5}.{\hat{6}} [\hat{6}\,\hat{K}]}
\nn\\
&=& \frac{i}{s_{12}}
 \frac{{\spb1.2}^3}{([2\,\hat{K}]\langle\hat{K}\,6\rangle)
                    (\langle6\,\hat{K}\rangle [\hat{K}\,1])}
 \frac{(\langle6\,\hat{K}\rangle [\hat{K}\,3])^3}
  {\spb3.4\spb4.5\spb{5}.{\hat{6}}([\hat{6}\,\hat{K}] \langle\hat{K}\,6\rangle)}
\,.
\label{NMHVkeq2}
\eea
Using \eqns{pppmmmid1}{pppmmmid3}, we can derive the identities,
\bea
\langle 6 \, \hat{K} \rangle \bigl[ \hat{K} \, a \bigr]
&=& \sandmm{6}.{(1+2)}.{a} \,, \nn\\
\spb{5}.{\hat{6}} &=& \spb5.6 + \frac{\spa1.2\spb5.1}{\spa6.2}
 = \frac{\sandmm{2}.{(6+1)}.{5}}{\spa6.2} \,, \nn\\
\bigl[ \hat{6} \, \hat{K} \bigr]
\langle \hat{K} \, 6 \rangle 
&=& \sandpp{6}.{(1+2)}.{6} + s_{12} = s_{612} \,,
\label{NMHV6ids}
\eea
where $s_{612} = (k_6+k_1+k_2)^2$.
Inserting these identities into \eqn{NMHVkeq2} for $T_2$, we have
\be
T_2 = i \, \frac{{\sandmm{6}.{(1+2)}.{3}}^3}
  {\spa6.1\spa1.2\spb3.4\spb4.5\,s_{612}\,\sandmm{2}.{(6+1)}.{5}} \,.
\label{T2final}
\ee

We can use the parity symmetry~(\ref{pppmmmparity}) to obtain the $k=4$
term.  The final result for the six-point NMHV amplitude is,
\bea
A_6(1^+,2^+,3^+,4^-,5^-,6^-) &=& 
 i \, \frac{{\sandmm{6}.{(1+2)}.{3}}^3}
  {\spa6.1\spa1.2\spb3.4\spb4.5\,s_{612}\,\sandmm{2}.{(6+1)}.{5}} \nn\\
&&\hskip-0.4cm\null
+  i \, \frac{{\sandmm{4}.{(5+6)}.{1}}^3}
  {\spa2.3\spa3.4\spb5.6\spb6.1\,s_{561}\,\sandmm{2}.{(6+1)}.{5}} \,.
\label{Apppmmm}
\eea
It's worth comparing the analytic form of this result to that found
in the 1980's~\cite{TreeColor2},
\bea
\label{Apppmmmold}
A_6(1^+,2^+,3^+,4^-,5^-,6^-) &=& 
  i \, \frac{(\spb1.2\spa4.5\sandmm{6}.{(1+2)}.{3})^2}
  {s_{61}s_{12}s_{34}s_{45}s_{612}} \\
&&\hskip-0.4cm\null
+ i \, \frac{(\spb2.3\spa5.6\sandmm{4}.{(2+3)}.{1})^2}
  {s_{23}s_{34}s_{56}s_{61}s_{561}} \nn\\
&&\hskip-0.4cm\null
+ i \, \frac{s_{123}\spb1.2\spb2.3\spa4.5\spa5.6
            \sandmm{6}.{(1+2)}.{3}\sandmm{4}.{(2+3)}.{1}}
  {s_{12}s_{23}s_{34}s_{45}s_{56}s_{61}} \,. \nn
\eea
Although the new form has only one fewer term, it represents the {\it physical}
singularities in a cleaner fashion.  For example, in the collinear limit
$3\parallel4$, \eqn{Apppmmm} makes manifest the $1/\spa3.4$ and $1/\spb3.4$
singularities, which correspond to the two different intermediate gluon
helicities that contribute in this collinear channel, as the six-point
NMHV amplitude factorizes on both the MHV and $\overline{\rm MHV}$
five-point amplitudes, $A_5(1^+,2^+,P^\pm,5^-,6^-)$. On the other hand,
each term of \eqn{Apppmmmold} behaves like the product of these two
singularities, since $1/s_{3,4} = -1/(\spa3.4\spb3.4)$.  Hence there
are large cancellations between the three terms in this channel.
Such cancellations can lead to large losses in numerical precision due to
round-off errors, especially in NLO calculations which typically evaluate
tree amplitudes repeatedly close to the collinear poles.

On the other hand, \eqn{Apppmmm} contains a {\it spurious} singularity
that \eqn{Apppmmmold} does not, as $\sandmm{2}.{(6+1)}.{5}\to0$.
This can happen, for example, whenever $k_6+k_1$ is a linear combination of
$k_2$ and $k_5$.  (In the collision $2+5\to 6+1+3+4$, such a configuration
is reached if the vectors $k_6+k_1$ and $k_3+k_4$ have no component transverse
to the beam axis defined by $k_2$ and $k_5$;
that is, if $k_6+k_1$ is a linear combintation of $k_2$ and $k_5$.)
It's called a spurious singularity because the amplitude should evaluate
to a finite number there, but individual terms blow up.  However, these
singularities tend to have milder consequences, as long as they appear
only to the first power, as they do here.
That's because the amplitude is not particularly large
in this region, so in the evaluation of an integral containing it by
importance-sampling, it is rare to come close enough to the surface where
$\sandmm{2}.{(6+1)}.{5}$ vanishes that round-off error is a problem.
Different choices of BCFW shifts lead to different spurious singularities,
so one can always check the value of $\sandmm{2}.{(6+1)}.{5}$ and use
a different shift if it is too small.

In general, the BCFW recursion relation leads to very
compact analytic representations for tree amplitudes.
The relative simplicity with respect
to previous analytic approaches becomes much more striking for seven or more
external legs.  A closely related set of recursion relations for ${\cal N}=4$
super-Yang-Mills theory~\cite{SuperBCFW} have been solved in closed form
for an arbitrary number of external legs~\cite{DrummondHennTrees}.
These solutions can also be used to compute efficiently a wide variety of
QCD tree amplitudes~\cite{DHPS}.  There are other ways to compute
tree amplitudes, in particular,
off-shell recursion relations based on the Dyson-Schwinger
equations, such as the Berends-Giele recursion relations~\cite{BGRecursion}.
At very high multiplicities, these can be numerically even more efficient
than the BCFW recursion relations.  Nevertheless, the idea behind the
BCFW recursion relations, that amplitudes can be reconstructed from their
analytic behavior, carries over to the loop level, as we'll now discuss.

\section{Generalized unitarity and loop amplitudes}
\label{sec:loops}

Ordinary unitarity is merely the statement that the scattering matrix $S$
is a unitary matrix, $S^\dagger S = 1$.  Usually we split off a
forward-scattering part by writing $S=1+iT$, leading to 
$(1-iT^\dagger)(1+iT)=1$, or
\be
{\rm Disc}\,T = T^\dagger T \,,
\label{ordinaryunitarity}
\ee
where ${\rm Disc}(x) = 2\,{\rm Im}(x)$ is the discontinuity across
a branch cut.
This equation can be expanded order-by-order in perturbation theory.
For example, the four- and five-gluon scattering amplitudes in QCD have
the expansions,
\bea
T_4 &=& g^2 T_4^{(0)} + g^4 T_4^{(1)} + g^6 T_4^{(2)} + \ldots,
\label{T4expansion}\\
T_5 &=& g^3 T_5^{(0)} + g^5 T_5^{(1)} + g^7 T_5^{(2)} + \ldots,
\label{T5expansion}
\eea
where $T_n^{(L)}$ is the $L$-loop $n$-gluon amplitude.
Inserting these expansions into \eqn{ordinaryunitarity} for the four-point
amplitude and collecting the coefficients at order $g^2$, $g^4$ and $g^6$,
respectively, we find that,
\bea
{\rm Disc}\,T_4^{(0)} &=& 0\,, \label{treenoIm}\\
{\rm Disc}\,T_4^{(1)} &=& T_4^{(0)\,\dagger} T_4^{(0)} \,, \label{oneloopIm}\\
{\rm Disc}\,T_4^{(2)} &=& T_4^{(0)\,\dagger} T_4^{(1)}
                        + T_4^{(1)\,\dagger} T_4^{(0)} 
                        + T_5^{(0)\,\dagger} T_5^{(0)} \,. \label{twoloopIm}
\eea
On the right-hand sides of these equations, there is an implicit discrete
sum over the types and helicities of the intermediate states which
lie between the two $T$ matrices, and there is a continuous integral over
the intermediate-state phase space.

The first equation (generalized to more legs) simply states that tree
amplitudes have no branch cuts.  The second equation, \eqn{oneloopIm},
states that the discontinuities of one-loop amplitudes are given by
the products of tree amplitudes, where the intermediate state always
consists of two particles that are re-scattering, the so-called
two-particle cuts.  The third equation, \eqn{twoloopIm},
states that the discontinuities of two-loop amplitudes are of two types:
two-particle cuts where one of the two amplitudes is a one-loop amplitude
rather than a tree amplitude, and three-particle cuts involving the product of
higher-multiplicity tree amplitudes.

Although there is a lot of information in \eqns{oneloopIm}{twoloopIm},
there are two more observations which lead to even more powerful conclusions.
The first observation is that the above unitarity relations are derived
assuming real momenta (and positive energies) for both the external states
and the intermediate states appearing on the right-hand sides.
The intermediate momenta on the right-hand sides can be thought of as
particular values of the loop momenta implicit on the left-hand side,
momenta that are real and on the particles' mass shell.
Given what we have learned so far about the utility of complex momenta at
tree level, it is natural to try to solve the on-shell conditions for
the loop momenta for complex momenta as well.  Such solutions are referred
to as {\it generalized} unitarity~\cite{ELOP}.

Secondly, because unitarity is being applied perturbatively, we might
as well make use of other the properties of perturbation theory,
namely that a Feynman diagram expansion exists.  We don't need to use
the actual values of the Feynman diagrams, but it is very useful to
know that such an expansion exists, because we can represent the loop
amplitudes as a linear combination of a basic set of Feynman
integrals, called {\it master} integrals, multiplied by coefficient
functions.  The idea of the {\it unitarity method}~\cite{Unitarity} is
that the information from (generalized) unitarity cuts can be compared
with the cuts of this linear combination, in order to determine all of
the coefficient functions.  If all possible integral coefficients can
be determined, then the amplitude itself is completely determined.
This approach avoids the need to use dispersion relations to
reconstruct full amplitudes from their branch cuts, which is often
necessary in the absence of a perturbative expansion.

In the rest of this section, we will sketch a useful hierarchical procedure
for determining one-loop amplitudes from generalized unitarity.
This method, and variations
of it, have been implemented both analytically, and even more powerfully,
numerically.  The latter implementation has made it possible to compute
efficiently one-loop QCD amplitudes of very high multiplicity, far beyond
what was imaginable a decade ago.  The availability of such loop amplitudes
has broken a bottleneck in NLO QCD computations, particularly for processes
at hadron colliders such as the LHC, leading to the ``NLO revolution.''

\subsection{The plastic loop integrand}

Before carrying out the loop integration, the integrand of a one-loop
amplitude depends on the external momenta $k_1,k_2,\ldots,k_n$ and
on the loop momentum $\ell$.  Just as at tree level, this function can
develop poles as the various momenta are continued analytically. Suppose
we hold the external momenta fixed and just vary $\ell$.
One kind of singularity that can appear is the ordinary two-particle cut
represented by \eqn{oneloopIm}.  Let's first generalize this equation to
the case of an $n$-gluon one-loop amplitude, and specialize it to the case
of a color-ordered loop amplitude $A_n^\oneloop$ --- the coefficient of the
leading-color single-trace color structure discussed in \sect{sec:color}.

\begin{figure}
\begin{center}
\includegraphics[width=5.5in]{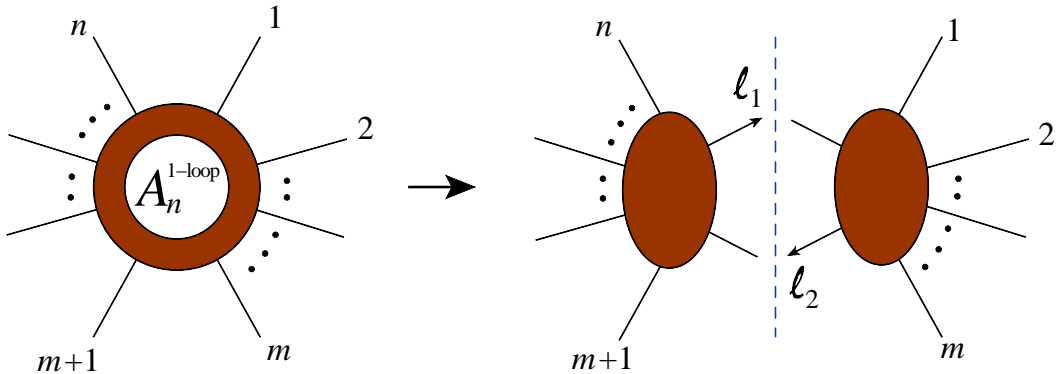}
\end{center}
\caption{Ordinary unitarity viewed as a factorization property of 
the loop integrand.}
\label{OrdUnitarityFigure}
\end{figure}

Consider the discontinuity in the channel $s_{12\ldots m}=(k_1+k_2+\cdots k_m)^2$,
which is illustrated in \fig{OrdUnitarityFigure}.
The unitarity relation that generalizes \eqn{oneloopIm} is
\bea
&&{\rm Disc}|_{s_{12\ldots m}} A_n^\oneloop(k_1,k_2,\ldots,k_n) \label{basiccut}\\
&=& (2\pi)^2 \! \sum_{h_i}\int \frac{d^D\ell_1}{(2\pi)^D}\;
\delta^{(+)}( \ell_1^\mu ) A_{m+2}^\tree(-\ell_1^{-h_1},k_1,\ldots,k_m,\ell_2^{h_2})\,,
\nn\\
&&\hskip2.6cm\null \times
 \delta^{(+)}(-\ell_2^\mu) A_{n-m+2}^\tree(-\ell_2^{-h_2},k_{m+1},\ldots,k_n,\ell_1^{h_1})
\eea
where $\ell_2 = \ell_1 - (k_1+k_2+\cdots k_m)$.  The delta function
$\delta^{(+)}(k^\mu) = \Theta(k^0) \delta(k^2)$ enforces that the
intermediate states are on shell with real momenta and positive energies.
The sum over intermediate helicities may also include different
particle types, for example, both gluons and quarks in an $n$-gluon QCD loop
amplitude.  The two delta functions reduce the loop momentum integral
to an integral over the two-body phase space for on-shell momenta
$\ell_1$ and $-\ell_2$.

Another way of stating \eqn{basiccut}, which allows us to generalize it,
is that for a given set of external momenta $k_i$, there is a family of
loop momenta $\ell\equiv\ell_1$ that solve the dual constraints
$\ell_1^2=\ell_2^2=0$.  On this solution set the loop integrand, which
can be pictured as the annular blob shown in \fig{OrdUnitarityFigure},
factorizes into the product of two tree amplitudes,
\be
\frac{i}{\ell_1^2}
A_{m+2}^\tree(-\ell_1^{-h_1},k_1,\ldots,k_m,\ell_2^{h_2})
\frac{i}{\ell_2^2}
A_{n-m+2}^\tree(-\ell_2^{-h_2},k_{m+1},\ldots,k_n,\ell_1^{h_1}) \,,
\label{twopartfact}
\ee
in much the same way that a tree amplitude factorizes on a single
multi-particle pole, \eqn{multiparticle}.

\begin{figure}
\begin{center}
\includegraphics[width=5.5in]{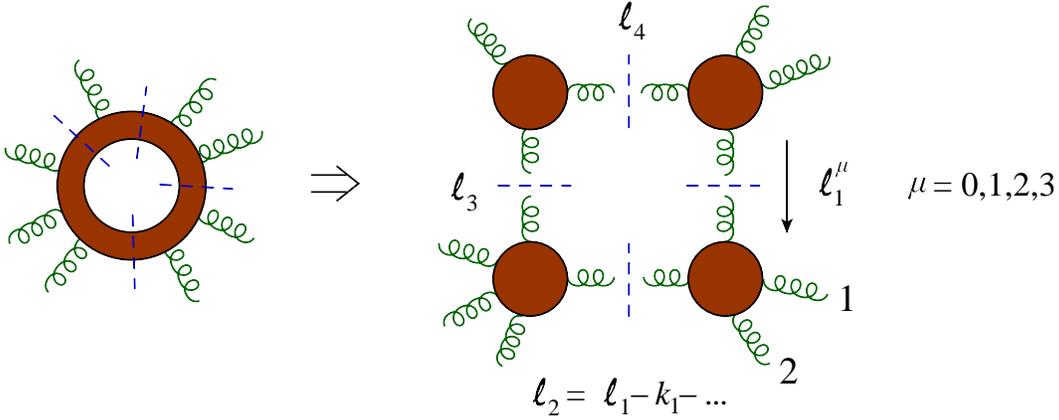}
\end{center}
\caption{A quadruple cut pinches the loop integrand down into the product
of four tree amplitudes, connected cyclicly around the loop.}
\label{QuadCutFigure}
\end{figure}

In this picture of the plastic loop integrand, we need not impose
positivity of the energies of the intermediate states, and the loop
momenta can even be complex.  This opens up the possibility of more general
solutions, where more than two lines are cut.  If we think of the loop
momentum $\ell^\mu$ as four-dimensional, then for generic kinematics we can
cut not just two lines, but up to four.  The reason the maximum is four
is that each cut
imposes a new equation of the form $(\ell-K_i)^2=0$ for some combination of
external momenta $K_i$.  At four cuts the number of equations equals
the number of unknowns --- the four components of $\ell^\mu$.
Hence a fifth cut condition is impossible to
satisfy (unless the kinematical configuration of the external momenta is
an exceptional, degenerate one).  \Fig{QuadCutFigure} shows how
the quadruple cut of a generic one-loop integrand squeezes it at four
locations, so that it becomes proportional to the product of four tree
amplitudes.  Two of the momenta of each tree amplitude are identified
with the cut loop momenta, denoted by $\ell_1,\ell_2,\ell_3,\ell_4$,
and the rest are drawn from the external momenta for the loop amplitude.


\subsection{The quadruple cut}

The quadruple cut~\cite{BCFGeneralized}
is special because the solution set is discrete.
Let's write the four cut loop momenta as
\be
\ell_1, \qquad 
\ell_2=\ell_1 - K_1, \qquad 
\ell_3=\ell_2 - K_2, \qquad 
\ell_4=\ell_3 - K_3 = \ell_1+K_4,
\label{ellidef}
\ee
where the $K_i$ are sums
of the $n$ external momenta satisfying $K_1+K_2+K_3+K_4=0$.
From \fig{QuadCutFigure} it is clear that the $K_i$ correspond to some
partition of the $n$ cyclicly ordered momenta into four contiguous sets.
We can rewrite the four quadratic cut conditions,
\be
\ell_1^2 = \ell_2^2 = \ell_3^2 = \ell_4^2 = 0,
\label{quadcutconditionA}
\ee
by taking the differences $\ell_i^2-\ell_{i+1}^2=0$,
so that three of the conditions are linear,
\be
\ell_1^2 = 0, 
\qquad 2\ell_1\cdot K_1 = K_1^2,
\qquad 2\ell_2\cdot K_2 = K_2^2,
\qquad 2\ell_3\cdot K_3 = K_3^2.
\label{quadcutconditionB}
\ee
Because the three linear equations can be solved uniquely,
we generically expect two discrete solutions for the loop momentum $\ell_1$, 
denoted by $\ell_1^\pm$.  The other three quantities
$\ell_i^\pm$ are uniquely determined from $\ell_1^\pm$ by shifting it by 
the appropriate external momenta.

\begin{figure}
\begin{center}
\includegraphics[width=4.5in]{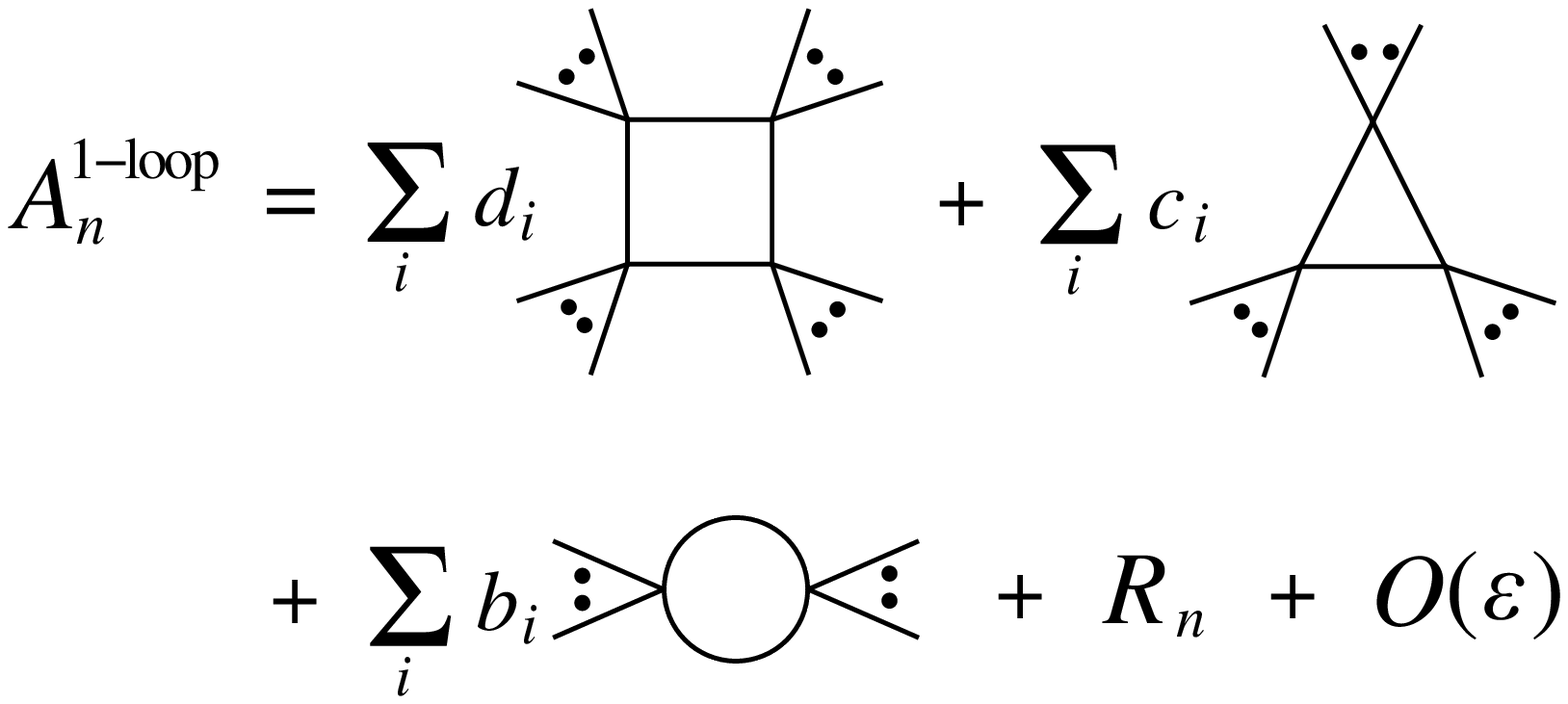}
\end{center}
\caption{Decomposition of a generic one-loop amplitude $A_n^\oneloop$
into basis integrals multiplied by kinematical coefficients:
scalar box integrals with coefficients $d_i$, scalar triangles
with coefficients $c_i$, scalar bubbles with coefficients $b_i$, and the
rational part $R_n$.  The dots between the external lines indicate that
one or several external legs may emanate from each vertex.
If there are massive internal propagators, then
tadpole integrals also appear; in the massless case such integrals vanish.}
\label{OneloopDecompFigure}
\end{figure}

What information does the quadruple cut reveal?  To answer this
question, we rely on a systematic decomposition of the one-loop
amplitude for an arbitrary $n$-point amplitude, which is shown
diagramatically in \fig{OneloopDecompFigure}.  The amplitude
can be written as a linear combination of certain {\it basis integrals},
multiplied by kinematical coefficients.  The only loop integrals that appear
are scalar integrals with four, three and two internal propagator lines, 
which are usually called box, triangle and bubble integrals, respectively.
They are given in dimensional regularization, with $D=4-2\e$, by
\bea
\I_4(K_1,K_2,K_3,K_4) &=&
\mu^{2\e} \int {d^{4-2\e}\ell \over (2\pi)^{4-2\e}}
{1 \over \ell^2 (\ell-K_1)^2 (\ell-K_1-K_2)^2 (\ell+K_4)^2} \,,
\label{I4def}\\
\I_3(K_1,K_2,K_3) &=&
\mu^{2\e} \int {d^{4-2\e}\ell \over (2\pi)^{4-2\e}}
{1 \over \ell^2 (\ell-K_1)^2 (\ell+K_3)^2} \,, \label{I3def}\\
\I_2(K) &=&
\mu^{2\e} \int {d^{4-2\e}\ell \over (2\pi)^{4-2\e}}
{1 \over \ell^2 (\ell-K)^2} \,, \label{I2def}
\eea
where the $K_i$ are the sums of external momenta emanating from
each corner.
The coefficients of these integrals are $d_i$, $c_i$ and $b_i$, where
$i$ labels all the inequivalent partitions of the $n$ external
momenta into 4, 3 and 2 sets, respectively.  There is also a rational
part $R_n$, which cannot be detected using cuts with four-dimensional cut
loop momenta; we will return to this contribution later.

The decomposition in \fig{OneloopDecompFigure}
holds in dimensional regularization, 
assuming that the external (observable) momenta are all four-dimensional,
and neglecting the $\Ord(\e)$ terms.  It also requires the internal
propagators to be massless; if there are internal propagators for
massive particles, then tadpole (one-propagator) integrals will also
appear.  The result seems remarkable at first sight, since
one-loop Feynman diagrams with five or more external legs attached to
the loop will generically appear, and these diagrams would seem
likely to generate pentagon and higher-point integrals.
However, it is possible to systematically reduce such integrals down
to linear combinations of scalar boxes, triangles and bubble
integrals~\cite{MelrosevNV,BDKPentagon,OtherPentagons}.

The reduction formulas are fairly technical, but here we don't need to
know the formulas, just that the reduction is possible.
Heuristically, the reason it is possible to avoid all pentagon
and higher-point integrals is the same reason that there is no quintuple cut
when the loop momentum is in four dimensions:
there are more equations in the quintuple cut conditions than there
are unknowns.  If the scalar pentagon integral had a quintuple cut, it
would not be possible to reduce it to a linear combination of box
integrals.  The fact that it can be done~\cite{MelrosevNV} exploits
the four-dimensionality of the loop momenta to expand the loop momenta
in terms of the four linearly-independent external momenta of the
pentagon.  In dimensional regularization, the relation of
ref.~\cite{MelrosevNV} has a correction term~\cite{BDKPentagon},
and the pentagon integral has a quintuple cut,
because the loop momentum is no longer four-dimensional.
However, because of the ``small'' volume of the extra $-2\e$ dimensions,
the correction term is of $\Ord(\e)$.

Returning to the quadruple cut, we see that a second special
feature of it is that only one of the integrals in
\fig{OneloopDecompFigure} survives, for a given quadruple cut.
First of all, none of the triangle and bubble terms can survive, because
those integrals do not even have four propagators available to cut.
There are many
possible box integrals, for a large number of external legs, but each one
box integral is in one-to-one correspondence with a different quadruple cut;
both are characterized by the same partition of the cyclicly ordered momenta
into four contiguous sets, or clusters.
The momentum flowing out at each corner of
the box must match the cluster momenta $\{K_1,K_2,K_3,K_4\}$ corresponding
to the quadruple cut~(\ref{quadcutconditionA}).
For this solution, we match the left- and right-hand sides of
\fig{OneloopDecompFigure} and learn~\cite{BCFGeneralized}
that
\be
d_i = \frac{1}{2} \Bigl( d_i^+ + d_i^- \Bigr),
\label{quadsolA}
\ee
where the superscripts $\pm$ refer to the two discrete solutions for the loop
momentum, and $d_i^\pm$ are given by the product of four tree
amplitudes, as in \fig{QuadCutFigure},
\be
d_i^\pm = A_1^\tree(\ell^\pm) A_2^\tree(\ell^\pm)
         A_3^\tree(\ell^\pm) A_4^\tree(\ell^\pm),
\label{quadsolB}
\ee
with
\be
A_i^\tree(\ell) \equiv A^\tree(-\ell_i,k^{(i)}_1,\ldots,k^{(i)}_{p_i},\ell_{i+1}).
\label{Aitree}
\ee
Here the external momenta $\{k^{(i)}_1,\ldots,k^{(i)}_{p_i}\}$ are the
elements of the cluster $K_i$, $i=1,2,3,4$,
{\it i.e.}~$\sum_{j=1}^{p_i} k^{(i)}_j = K_i$.
These formulae are very easy to evaluate, either analytically or
in an automated code, and they are numerically very stable.

It's possible to solve analytically for the cut loop momenta $\ell_i^\pm$
for generic values of the $K_i$; the solution involves a quadratic
formula~\cite{BCFGeneralized}.  If just one of the external legs
is massless, however, say $K_1 = k_1$, then the solutions collapse
to a simpler form~\cite{BlackHat,RisagerThesis}:
\begin{eqnarray}
&&(\ell_1^\pm)^\mu =  {\sandmppm1.{\s K_2 \s K_3 \s K_4 \gamma^\mu}.1\over
  2 \sandmppm1.{\s K_2 \s K_4}.1} \,,
\hskip 2cm 
(\ell_2^\pm)^\mu = 
- {\sandmppm1.{\gamma^\mu \s K_2 \s K_3 \s K_4}.1 \over  
  2 \sandmppm1.{\s K_2 \s K_4}.1} \,,\nn \\
&&(\ell_3^\pm)^\mu =  
{\sandmppm1.{\s K_2 \gamma^\mu  \s K_3 \s K_4}.1 \over
  2 \sandmppm1.{\s K_2 \s K_4}.1} \,,
\hskip 2cm 
(\ell_4^\pm)^\mu = 
- {\sandmppm1.{\s K_2 \s K_3 \gamma^\mu  \s K_4}.1\over
  2 \sandmppm1.{\s K_2 \s K_4}.1} \,. \hskip 1 cm 
\label{MasslessSolution}
\end{eqnarray}
It's easy to see that \eqn{quadcutconditionA} is satisfied
by \eqn{MasslessSolution}; that is, each of the four vectors 
$(\ell_i^\pm)^\mu$ squares to zero.  For example,
the evaluation of $(\ell_1^\pm)^\mu(\ell_1^\pm)_\mu$ proceeds using the Fierz
identity and is proportional to $\spa1.1 = 0$.  The corresponding
algebra for $(\ell_3^\pm)^2$ involves
$\sandmppm1.{\s K_2 \s K_2}.1 = K_2^2 \spa1.1 = 0$.

We also have to show that momentum conservation is satisfied, namely,
\be
\ell_2 - \ell_3 = K_2 \,, \qquad
\ell_3 - \ell_4 = K_3 \,, \qquad
\ell_4 - \ell_1 = K_4 \,.
\label{quadcutmomcons}
\ee
The first equation is
\be
(\ell_2^\pm - \ell_3^\pm)^\mu = 
- {\sandmppm1.{ \{ \gamma^\mu, \s K_2 \} \s K_3 \s K_4}.1 \over  
  2 \sandmppm1.{\s K_2 \s K_4}.1}
= - K_2^\mu {\sandmppm1.{ ( - \s k_1 - \s K_2 - \s K_4 ) \s K_4}.1 \over  
  \sandmppm1.{\s K_2 \s K_4}.1}
= K_2^\mu
 \,, \label{quadcutmc1}
\ee
and the other equations work the same way.

\begin{figure}
\begin{center}
\includegraphics[width=1.8in]{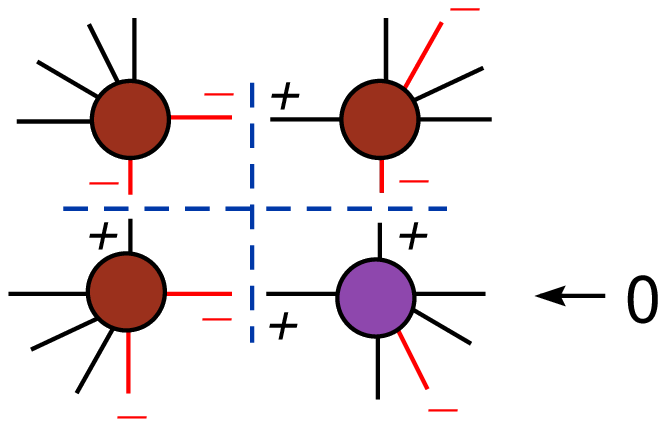}\hskip1.4cm
\includegraphics[width=2.6in]{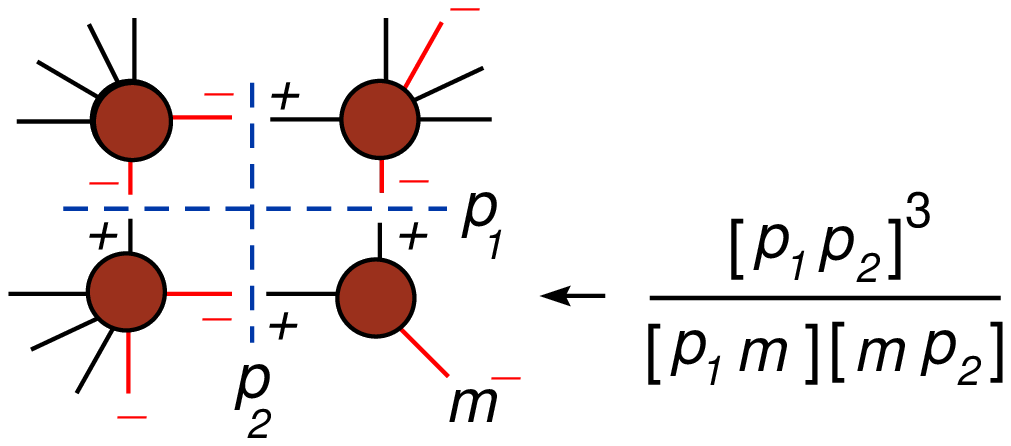}
\end{center}
\caption{The left quadruple cut shows that the coefficients
of all four-mass box integrals vanish for one-loop NMHV amplitudes.
The right quadruple cut shows that the three-mass box coefficients
do not vanish.}
\label{QuadCutVanishFigure}
\end{figure}

Shortly, we will compute an explicit example of a nontrivial, nonzero
coefficient of a box integral using the quadruple cut.
However, it's worth noting first that many box coefficients for massless
QCD amplitudes vanish identically.
In fact, the vanishing of large sets of box coefficients can be
established simply by counting negative helicities.
Consider, for example, the one-loop NMHV amplitude in massless QCD
whose quadruple
cut is shown on the left side of \fig{QuadCutVanishFigure}.
This quadruple cut can be used to compute the coefficient of
a four-mass box integral.  We call it a four-mass box because the
momentum $K_i$ flowing out at each corner is the sum of at least two
massless external particle momenta; hence $K_i$ is a massive four-vector.
(In contrast, the right side of \fig{QuadCutVanishFigure} shows a quadruple
cut for a three-mass box integral, because the lower right tree amplitude
emits a single external momentum $m$.)

We denote negative-helicity legs by red lines and an explicit $(-)$
in the figure.  The external black lines are all positive helicity.
The upper left tree amplitude in the example has no external negative
helicities.  Because tree amplitudes with 0 or 1 negative helicity vanish,
according to \eqn{treevanish}, the two internal (cut) lines emanating
from this upper left blob must carry negative helicity.  On the opposite
side of their respective cuts, they carry positive helicity.
If the lower left and upper right tree amplitudes have one negative
external helicity, as shown, then they must each send a negative helicity
state toward the purple blob.  This tree amplitude carries the third
external negative helicity, but no other negative helicity emanates from it,
so it vanishes, causing the vanishing of the corresponding four-mass
box coefficient.

We gave this argument specifically for the case
that all three negative-helicity particles were emitted from different
corners of the box.  It's easy to see that the vanishing does not
actually depend on where the negative helicities are located.
It's simply a reflection of the
fact that there are four tree amplitudes, all with more than three legs,
so there must be at least $4\times2=8$ negative helicities among the
external and cut legs.  However, each cut has exactly one negative helicity,
and there are three negative external helicities, for a total of $4+3 = 7$.
Since $7<8$, the NMHV four-mass box coefficients always vanish.
This counting argument fails as soon as one of the corner momenta
becomes massless, as is appropriate for the three-mass cut shown
on the right side of \fig{QuadCutVanishFigure}.  With the right
(second) type of complex kinematics discussed in \sect{sec:kin3pt},
the three-point tree amplitude with helicity configuration $({+}{+}{-})$
is nonvanishing, as shown in the figure.  Hence this three-mass
box coefficient is nonvanishing.  There is a single
quadruple-cut helicity configuration and a single choice
of sign for the kinematical configuration~(\ref{MasslessSolution})
that contributes in the particular case shown.

Using the same counting argument, we can see that one-loop
MHV amplitudes, with two external negative helicities, contain
neither four-mass, nor three-mass, box integrals.
The two-mass box integrals can be divided into two types,
``easy'', in which the two massive corners are diagonally opposite,
and ``hard'', in which they are adjacent to each other.
One can show that the hard two-mass boxes always vanish as well.
(This proof can be done with the help of a triple cut which puts
the two massless corners into one of the three trees.  Then the counting
of negative helicities is analogous to the four-mass NMHV example,
except that one needs $3\times2=6$ negative helicities, and one has
only $3+2=5$ available.)

As an aside, consider the one-loop amplitudes of the form
$A_n^\oneloop(1^\pm,2^+,\ldots,n^+)$, for which the corresponding
tree amplitudes vanished according to \eqn{treevanish}.
A similar counting exercise shows that they have no cuts
at all:  no quadruple, triple, or ordinary two-particle cuts.
They are nonvanishing (at least in a non-supersymmetric theory like QCD),
but they are forced to be purely rational functions of the external
kinematics~\cite{LastFinite}.


\subsection{A five-point MHV box example}

\begin{figure}
\begin{center}
\includegraphics[width=2.3in]{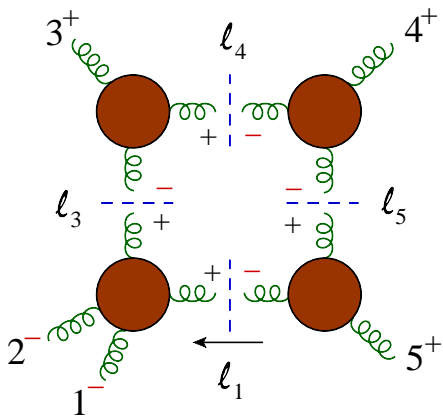}
\end{center}
\caption{The quadruple cut for one of the box coefficients
for the five-gluon amplitude with helicity configuration $({-}{-}{+}{+}{+})$.}
\label{QuadCutmmpppFigure}
\end{figure}

In the remainder of this section, we will compute one of the box
coefficients for the five-gluon QCD amplitude
$A_5^\oneloop(1^-,2^-,3^+,4^+,5^+)$, the one in which the two negative
helicity legs, 1 and 2, are clustered into a massive
leg (as also reviewed in ref.~\cite{OnShellReview}).
The quadruple cut for this box coefficient is shown
in \fig{QuadCutmmpppFigure}.  Inspecting the figure,
starting with the lower-left tree amplitude, it is clear that there
is a unique assignment of internal helicities.  Also, this
assignment of helicities forbids quarks (or scalars) from propagating
in the loop; the tree amplitudes for two spin $1/2$ fermions (or two scalars)
and two identical helicity gluons vanish (see \eqn{treeqgvanish} for
the fermion case).  Therefore this box coefficient receives contributions
only from the gluon loop, and is the same in QCD as in gauge theories
with different matter content (such as ${\cal N}=4$ super-Yang-Mills theory).

Now that we have identified which four tree amplitudes are to be
multiplied together, the next task is to determine the cut loop momentum.
In particular, let's work out $\ell_4$, the loop momentum just before
the massless external leg 4.  We can use \eqn{MasslessSolution}, but 
since leg 1 was massless there, we should relabel the momenta in
that equation according to:
\be
\ell_1^\pm \to \ell_4^\pm \,, \quad
k_1 \to k_4 \,, \quad
K_2 \to k_5 \,, \quad
K_3 \to k_1 + k_2 \,, \quad
K_4 \to k_3 \,.
\label{QCmmpppRelabel}
\ee
Then the first equation in (\ref{MasslessSolution}) becomes,
\be
(\ell_4^\pm)^\mu 
=  {\sandmppm4.{5 (1+2) 3 \gamma^\mu}.4\over 2 \sandmppm4.{5 3}.4}
= - {\sandmppm4.{5 4 3 \gamma^\mu}.4\over 2 \sandmppm4.{5 3}.4}
= - {\sandpmpm5.{4 3 \gamma^\mu}.4\over 2 \sandpmpm5.3.4} \,.
\label{ell4A}
\ee
Which sign should we use?  The sign is dictated by the helicity assignments
in the three-point amplitudes.  Because the upper-right tree
is of type $({-}{-}{+})$, and is constructed from right-handed spinors,
the three left-handed spinors should be proportional.
In particular, $\tilde\lambda_{\ell_4} \propto \tilde\lambda_{4}$, which
tells us that we should take the lower sign in \eqn{ell4A},
so that
\be
\ell_4^\mu = (\ell_4^-)^\mu 
= \frac{1}{2} \frac{\spa4.5}{\spa3.5} \spab3.{\gamma^\mu}.4 \,.
\label{ell4B}
\ee

Now we can multiply together the four tree amplitudes, and use
\eqns{quadsolA}{quadsolB} (with $d_i^+=0$) to get for the ``$(12)$'' box
coefficient,
\bea
d_{(12)} &=&
\frac12 A_4^\tree(-\ell_1^+,1^-,2^-,\ell_3^+)
A_3^\tree(-\ell_3^-, 3^+, \ell_4^+)
A_3^\tree(-\ell_4^-,4^+,\ell_5^-)
A_3^\tree(-\ell_5^+,5^+,\ell_1^-)
\nn\\
&=& 
\frac12
\frac{\spa1.2^3}{\spa2.{\ell_3}\spa{\ell_3}.{(-\ell_1)}\spa{(-\ell_1)}.1}
\frac{\spb3.{\ell_4}^3}{\spb{\ell_4}.{(-\ell_3)}\spb{(-\ell_3)}.3}
\frac{\spa{\ell_5}.{(-\ell_4)}^3}{\spa4.{\ell_5}\spa{(-\ell_4)}.4}
\frac{\spb{(-\ell_5)}.5^3}{\spb5.{\ell_1}\spb{\ell_1}.{(-\ell_5)}}
\nn\\
&=&
 -\frac12\frac{\spa1.2^3 \sandpm3.{\Ellin_{4}\Ellin_{5}}.5^3}
{\sand2.{\Ellin_{3}}.3\sand4.{\Ellin_{4}\Ellin_{3}\Ellin_{1}}.5
\sandmp1.{\Ellin_{1}\Ellin_{5}}.4}
\,.
\label{d12A}
\eea
To get to the last step in \eqn{d12A}, we combined spinor products into
longer strings using the replacement $|\ell_i\rangle [\ell_i| \to \s{\ell_i}$,
but we did not need to use any other properties of the $\ell_i$.
In the next step it is convenient to use momentum conservation,
{\it i.e.}~$\ell_1=\ell_4-k_4-k_5$, $\ell_3=\ell_4+k_3$
and $\ell_5=\ell_4-k_4$, as well as $\ell_i^2=0$, to replace,
\bea
\sandpm3.{\Ellin_{4}\Ellin_{5}}.5 &\to& - \spab4.{\ell_4}.3 \spa4.5 \,,\\
\sandmm2.{\Ellin_{3}}.3 &\to& \sandmm2.{\Ellin_{4}}.3 \,,\\
\sand4.{\Ellin_{4}\Ellin_{3}\Ellin_{1}}.5 &\to&
 \sand4.{\Ellin_{4}3(\Ellin_{4}-k_4)}.5
 = - \sand4.{\Ellin_{4}}.3 \spa3.4 \spb4.5
\label{mideq} \,,\\
\sandmp1.{\Ellin_{1}\Ellin_{5}}.4 &\to& - \spa1.5 \spab4.{\ell_4}.5 \,.
\eea
In \eqn{mideq} we also used the fact that $\spa3.{\ell_4}=0$, given that
both $\ell_4$ and $k_3$ emanate from a $({+}{+}{-})$ three-point
amplitude.

Making these replacements in \eqn{d12A}, and then Fierzing in
$\ell_4^\mu \propto \spab3.{\gamma^\mu}.4$ from \eqn{ell4B}, gives,
\bea
d_{(12)} &=&
 \frac12\frac{\spa1.2^3\sand4.{\Ellin_{4}}.3^2\spb4.5^3}{\sand2.{\Ellin_{4}}.3
\spa3.4\spb4.5\spa1.5\sand4.{\Ellin_{4}}.5}
\nn\\
&=&
 -\frac12\frac{\spa1.2^3 s_{34} s_{45}}{\spa2.3\spa3.4\spa4.5\spa5.1}
\nn\\
&=&
\frac{i}2 s_{34} s_{45} \, A_5^\tree(1^-,2^-,3^+,4^+,5^+)\,.
\label{d12B}
\end{eqnarray}

For completeness, we give the formula for the one-mass box integral
multiplying this coefficient.  It is defined in \eqn{I4def} and
has the Laurent expansion in $\e$,
\bea
\I_4^{(12)}
&=& { - 2 i \, \cg \over s_{34} s_{45} } \biggl\{
-{1\over\e^2} \biggl[ 
    \biggl({\mu^2\over -s_{34}}\biggr)^{\e}
     +\biggl({\mu^2\over -s_{45}}\biggr)^{\e}
     -\biggl({\mu^2\over -s_{12}}\biggr)^{\e}
    \biggr]
\nn\\
&&\hskip1.2cm
  + \Li_2\biggl(1-{s_{12}\over s_{34}}\biggr)
  + \Li_2\biggl(1-{s_{12}\over s_{45}}\biggr)
  + {1\over2} \ln^2\biggl({-s_{34}\over -s_{45}}\biggr)
  + {\pi^2\over6}
\biggr\} 
\nn\\
&&\hskip0.0cm
+\ \Ord(\e) \,,
\eea
where the constant $\cg$ is defined by
\be
\cg = {1\over(4\pi)^{2-\e}}
  {\Gamma(1+\e)\Gamma^2(1-\e)\over\Gamma(1-2\e)} \,.
\label{cgdefn}
\ee

Interestingly, the result~(\ref{d12B}) is proportional to the tree
amplitude.  The coefficients of the four other box integrals (labeled
$(23)$, $(34)$, $(45)$ and $(51)$) also have only gluonic contributions
for this helicity choice, and their coefficients turn out to be given by
cyclic permutations of \eqn{d12B}.  Hence we have for the gluonic
contribution to the one-loop amplitude,
\begin{eqnarray}
\hskip-0.5cm
A_5^\oneloop(1^-,2^-,3^+,4^+,5^+)
&=& A_5^\tree(1^-,2^-,3^+,4^+,5^+) \, \cg \, \biggl\{
\nonumber\\
&&\hskip0.0cm
-{1\over\e^2} \biggl[ 
    \biggl({\mu^2\over -s_{34}}\biggr)^{\e}
     +\biggl({\mu^2\over -s_{45}}\biggr)^{\e}
     -\biggl({\mu^2\over -s_{12}}\biggr)^{\e}
    \biggr]
\nonumber\\
&&\hskip0.0cm
  + \Li_2\biggl(1-{s_{12}\over s_{34}}\biggr)
  + \Li_2\biggl(1-{s_{12}\over s_{45}}\biggr)
  + {1\over2} \ln^2\biggl({-s_{34}\over -s_{45}}\biggr)
  + {\pi^2\over6}
\nonumber\\
&&\hskip0.0cm
+\ \hbox{cyclic permutations} \biggr\}
\nonumber\\
&&\hskip0.0cm
+\ \hbox{triangles + bubbles + rational}.
\end{eqnarray}

If we were computing the amplitude in ${\cal N}=4$ super-Yang-Mills
theory, we would be done at this point:  One can show that the
triangles, bubbles and rational parts all vanish in this
theory~\cite{Unitarity}.  In the case of QCD, there is more work to do.
In the next subsection we sketch a method~\cite{Forde,BlackHat}
for determining the triangle coefficients.


\subsection{Triangle coefficients}

\begin{figure}
\begin{center}
\includegraphics[width=5in]{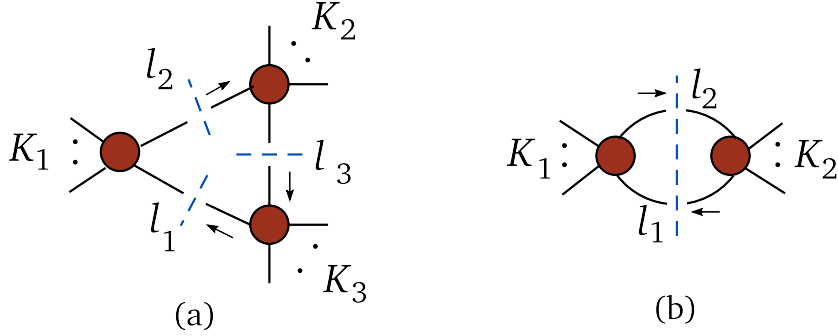}
\vskip -.3 cm 
 \caption{(a) The triple cut and (b) the ordinary double cut used to
determine the coefficients of the triangle and bubble integrals.  The
loop momenta $l_i$ are constrained to satisfy on-shell conditions.}
\label{TripleDoubleCutFigure}
\end{center}
\end{figure}

By analogy, we expect the triangle coefficients to be determined by the
triple cut shown in \fig{TripleDoubleCutFigure}(a), and the bubble
coefficients by the double cut shown in \fig{TripleDoubleCutFigure}(b).
The solution to the three equations defining the triple cut,
\be
\ell_1^2(t) = \ell_2^2(t) = \ell_3^2(t) = 0,
\label{triplecutcondition}
\ee
depends on a single complex parameter $t$.
However, the triple cut generically also receives contributions
from the box integral terms in \fig{OneloopDecompFigure}.
The box contributions have to be removed before identifying the
coefficient of a given scalar triangle integral.
Take any one of the three tree amplitudes in
\fig{TripleDoubleCutFigure}(a), and imagine pinching that blob until
it splits into two, exposing another loop propagator.  This corner
of the triple-cut phase space has the form of
a box integral contribution.  The pinching imposed a fourth cut
condition, which has discrete solutions, so it must only occur
at discrete values of $t$, say $t_i^\sigma$ where $i$ labels the different
quadruple cuts that sit ``above'' the given triple cut, and $\sigma=\pm$
labels the two possible discrete solutions.

The generic form of the triple cut is
\bea
C_3(t) &=& A^\tree_{(1)}(-\ell_1,k_1,\ldots,k_{p_1},\ell_2)
           A^\tree_{(2)}(-\ell_2,k_{p_1+1},\ldots,k_{p_2},\ell_3)
           A^\tree_{(3)}(-\ell_3,k_{p_2+1},\ldots,k_{n},\ell_1) \nn\\
 &=& T_3(t) + \sum_{\sigma=\pm} \sum_i 
\frac{d_i^\sigma}{\xi_i^\sigma (t-t_i^\sigma)} \,,
\label{CTdecomp}
\eea
where $d_i^\sigma$ are the previously computed box
coefficients~(\ref{quadsolB}), and $T_3(t)$ is the triple cut
``cleaned'' of all singularities at finite $t$.

The pole locations $t_i^\sigma$ and the residue factors $\xi_i^\sigma$
do not depend on the amplitude being calculated, but only on
the kinematics of the relevant triple and quadruple cuts.
They can be computed from the solution for $\ell_i(t)$.
For massless internal particles, the solution of \eqn{triplecutcondition}
is~\cite{AguilaPittau,OPP,Forde}
\be
\ell^{\mu}_1(t) = \Kfmu1 +
\Kfmu3 +
\frac{t}{2}\Bmp{\Kfm1|\gamma^{\mu}|\Kfm3}
+\frac{1}{2t}
\Bmp{\Kfm3|\gamma^{\mu}|\Kfm1}\,,
\label{TripleCutMomentum}
\ee
and, using momentum conservation, $\ell_2(t)=\ell_1(t)-K_1$, 
$\ell_3(t)=\ell_1(t)+K_3$. 
Here we have introduced a pair of massless auxiliary vectors
$\Kfmu{1}$ and $\Kfmu{3}$, constructed from $K_1$ and $K_3$,
\begin{eqnarray}
\Kfmu1=\gamma\alpha\frac{\gamma K_1^{\mu}+S_1 K^{\mu}_3}{\gamma^2-S_1S_3}\,,
\hskip 2 cm  
\Kfmu3=-\gamma\alpha'\frac{\gamma K_3^{\mu}+S_3 K_1^{\mu}}{\gamma^2-S_1S_3} \,,
\label{eq:def_gamma_pm}
\end{eqnarray} 
where $S_1 = K_1^2$, $S_3 = K_3^2$, and
\begin{eqnarray}
\alpha = \frac{S_3(S_1-\gamma)}{S_1S_3-\gamma^2}\,,\qquad
\alpha' = \frac{S_1(S_3-\gamma)}{S_1S_3-\gamma^2}\,,\qquad
\gamma = \gamma_{\pm} = - K_1\cdot K_3 \pm\sqrt{\Delta}\,,
\end{eqnarray} 
with
\begin{equation}
\Delta = (K_1\cdot K_3)^2 - K_1^2K_3^2 \,.
\end{equation}

The coefficient of the scalar triangle integral is the ``$t$ independent''
part of the triple cut.  To be more precise, the quantity $T_3(t)$
has no singularities at finite values of $t$ because they are all accounted
for by the box contributions shown explicitly in \eqn{CTdecomp}.
Because this quantity has singularities only at $t=0$ and $t=\infty$,
it can be represented as,
\be
T_3(t) = \sum_{k=-p}^p c_k t^k \,.
\label{T3Fourier}
\ee
The desired quantity, the triangle coefficient, is $c_0$.  The other terms
correspond to tensor triangle integrals that integrate to zero
(``spurious terms'' in the language of OPP~\cite{OPP}).
For renormalizable theories, there are at most three loop momenta
in the numerator of triangle integrals, and one can take $p=3$.

One way to isolate $c_0$ is from the $t^0$ term in the large $t$ limit
of $T_3(t)$, or of $C_3(t)$ itself, since the box contributions go to zero
in this limit.  This is an effective method for determining $c_0$
analytically~\cite{Forde}.   For an automated implementation, the $t^0$
term is usually subleading as $t\to\infty$, making it difficult to
extract numerically.  Instead one can work at finite $t$, and
extract $c_0$ (and the other $c_k$ coefficients) out of the finite sum
in \eqn{T3Fourier} by using the discrete Fourier
projection,
\begin{equation}
c_k = {1\over 2p+1} \sum_{j = -p}^{p} \,
         \Bigl[ t_0 e^{2 \pi i j/(2p + 1)} \Bigr]^{-k} \, 
   T_3\Bigl(t_0 e^{2 \pi i j/(2p + 1)}\Bigr)\,,
\label{ckeqn}
\end{equation}
for some choice of $t_0$.
This approach is very stable numerically~\cite{BlackHat}.

The other $c_k$ coefficients are actually needed in the next step,
the determination of the bubble coefficients.  The double cut
depends on two complex parameters.  It has singularities corresponding
to both triple cuts and quadruple cuts, which can be ``cleaned''
in a fashion analogous to \eqn{CTdecomp}, using the previously
computed box and triangle information.  Because the triple cut
depends on a complex parameter, all of the $c_k$ coefficients are required
to characterize it.  After cleaning the double cut, a double discrete sum
analogous to \eqn{ckeqn}
can be used to extract the bubble coefficient.  For real cut momenta,
the two parameters of the double cut have a simple physical interpretation:
they are just the angles $\theta,\phi$ of one of the two intermediate
states, in the center of mass frame for the channel being cut.
The double discrete sum essentially performs a spherical harmonic
expansion (it is slightly different because the intermediate
momenta can be treated as complex).

The hierarchical determination of the ``cut-constructible'' parts
of one-loop amplitudes described here~\cite{BlackHat} is quite similar to the
OPP method~\cite{OPP} and to the method described in ref.~\cite{EGK}, all
of which have been implemented in an automated fashion.


\subsection{The rational part}

The last remaining part of the amplitude is the rational part $R_n$.
This component cannot be detected by any unitarity
cut in which the cut loop momentum are confined to four dimensions.
We have implicitly been assuming throughout this section that the $\ell_i$
are four-dimensional.  This assumption was very convenient because it allowed
us to label the states with four-dimensional helicities, and use all
the vanishing relations for the tree amplitudes that enter the
four-dimensional cuts.  One way to determine the rational
part, called $D$-dimensional
unitarity~\cite{DDimUnitarity,AnnRev,DDimUnitarityA}, is to let the cut momenta
have extra-dimensional components, thinking of the $\e$ in $D=4-2\e$ as a
negative number.  In this approach, there are also nonvanishing quintuple
cuts.  There are no hexagon cuts because at one loop, all extra-dimensional
components of the loop momentum are equivalent; they might as well point
in a single, fifth dimension.  So there are five components of the loop
momentum that can be constrained by generalized cuts.
The same kind of hierarchical, automated approach described above
can be applied to the $D$-dimensional case~\cite{DDimUnitarityAuto}.
In this case, one does not need to determine every extra-dimensional
term in the loop integrand; the measure factor is $d^{-2\e}\ell$,
leading to an integral of $\Ord(\e)$, unless there are enough
factors of the extra-dimensional components, denoted by
$\ell_{(-2\e)}^2\equiv \mu^2$,
in the numerator of the loop integrand
to generate a compensating factor of $1/\e$.
For more details on this method, see the review~\cite{EKMZReview}.

A second method for computing the rational part is to apply a BCFW
shift to the integrated loop amplitude.  This approach can be implemented
analytically~\cite{OnShellReview}, and numerically~\cite{BlackHat}.
Here we just mention a few salient points.
When a complex $z$-dependent shift is applied to a tree amplitude,
as in \sect{sec:trees}, the result is a meromorphic function of $z$,
where the poles correspond to factorization of the tree amplitude
into two lower-point amplitudes.  When the same shift is applied to a loop
amplitude, branch cuts in $z$ are generated, from the logarithms and
dilogarithms appearing in the scalar integrals.  There are also poles,
whose origin from amplitude factorization is similar to the tree-level case.
The branch cuts would complicate an analysis of the poles.
However, if we have already computed the cut part $C_n$, we can consider
shifting only the rational part, $R_n = A_n - C_n \to R_n(z)$.

The function $R_n(z)$ is meromorphic, so we can contemplate computing
$R_n(0)$ from Cauchy's theorem, using an equation analogous to \eqn{Cauchy1},
if we know all of its poles and residues.  However, $R_n(z)$ has two different
types of poles.  The {\it physical} poles are the ones that appear in
$A_n(z)$, and their residues can be computed from factorization in
a similar fashion to tree level.  There is a second set of {\it spurious}
poles.  These poles are not poles of $A_n(z)$.  They come from singularities
in kinematical regions where $A_n$ is non-singular, but $C_n$ and $R_n$
separately diverge.  (One example of such a region is where 
$\sandmm{2}.{(6+1)}.{5}\to0$; see \sect{sec:BCFWNMHVapp}.)
Becasue $A_n(z)$ has no poles, the spurious-pole residues in $R_n(z)$
must be the negative those in $C_n(z)$.  Because the cut part is known and
the locations of all the spurious poles are known, the residues of $C_n(z)$
are straightforward to compute.  For more details on this method, see the
review~\cite{OnShellReview}.

Within the OPP method~\cite{OPP}, the rational part is given by a sum of
two terms, called $R_1$ and $R_2$.  The $R_1$ part is obtained as a byproduct
of the computation of the cut part, by taking into account the
extra-dimensional $\mu^2$ dependence appearing in the propagator denominators
of the dimensionally-regulated loop integrand~\cite{OPP2}.
The remaining $R_2$ terms come from $\mu^2$ dependence in
the numerator of the loop integrand.  As in the $D$-dimensional unitarity
method, only a limited set of terms have enough factors of $\mu^2$
in the numerator to produce a nonzero rational term.  For renormalizable
theories, these contributions can be computed for all processes, in
terms of a relatively small number of effective two-, three- and four-point
vertices~\cite{OPP2,OPPrat}.

These new, efficient methods have enabled the construction of a variety
of automated computer programs for generating one-loop amplitudes, including
{\sc CutTools}~\cite{CutTools},
{\sc BlackHat}~\cite{BlackHat}, {\sc Rocket}~\cite{Rocket},
{\sc SAMURAI}~\cite{SAMURAI}, {\sc NGluon}~\cite{NGluon},
{\sc MadLoop}~\cite{MadLoop}, {\sc HELAC-NLO}~\cite{HELACNLO},
{\sc GoSam}~\cite{GoSam}, {\sc Open Loops}~\cite{OpenLoops} and
{\sc Recola}~\cite{Recola}.

For NLO QCD corrections to collider processes, it is also necessary to
consider tree-level processes with one additional parton radiated into
the final state, and integrate their cross section over a phase space
that contains the soft and collinear singularities discussed in
\sect{sec:factorization}.  A variety of efficient, automated methods
have been developed recently for performing these phase-space
integrals, based on the methods originally developed in
refs.~\cite{FKS,CS}.  In combination with the one-loop methods
sketched here, these methods have led to a variety of NLO QCD results
for LHC processes with four, five and even six objects (electroweak
particles or jets) in the final state.  They have opened up a new
avenue for precision theory at hadron colliders, which has proved to
be very important for gaining quantitative control over important
Standard Model backgrounds, as well as for performing detailed
experimental studies of QCD dynamics.

\section{Conclusions}
\label{sec:concl}

In these notes, we have only scratched the surface of modern techniques for
computing scattering amplitudes.  We covered the general formalism
and factorization properties of helicity amplitudes, explored tree-level
analyticity and the BCFW recursion relation, and described some of the
techniques for using generalized unitarity at one loop.  Numerous additional
details are required in order to assemble full one loop QCD amplitudes,
many of which are discussed in other
reviews~\cite{OnShellReview,BrittoReview,ItaReview}, and
in particular the comprehensive review~\cite{EKMZReview}.

We did not touch on multi-loop scattering amplitudes at all, but this
is an exceedingly rich subject.  Amplitudes in ${\cal N}=4$ super-Yang-Mills
theory --- QCD's maximally supersymmetric cousin ---
have been computed using similar ideas, through many loops and for
many external legs.  Remarkable properties have been found, leading
to new approaches.  For more in this direction, as well as applications
to supergravity, the reader can consult the very recent, authoritative
review~\cite{EHReview}.  

The multi-loop applications of unitarity-based
methods to QCD are still in their infancy, but they are being developed
very rapidly now.  For the simplest $2\to2$ processes, the
principles of generalized unitarity were applied a while
ago~\cite{AllPlus2,Ourgggg2,TwoloopSplit},
but not in a way that could be automatically extended
to more complicated processes.
The latter direction has seen important recent
progress~\cite{JKL2loops,BFZFivegluon2,MMOT,KMPV}, 
but there is still a ways to go before
two-loop QCD amplitudes for generic $2\to3$ processes will be available.
A large part of the problem is not just determining the loop integrand,
but evaluating all the loop integrals.

I hope that some of you who have made it this far will be encouraged
to explore further, and indeed to push the boundaries of our knowledge
about scattering amplitudes and their applications to collider physics
as well as other problems.

\section*{Acknowledgments}

I am grateful to Christophe Grojean, Martin Mulders, Maria Spiropolu,
Bogdan Dobrescu and Iain Stewart for the invitations to give these 
two sets of lectures and for the encouragement to prepare these notes.
I also thank the students at both schools for their enthusiasm and
excellent questions.  Finally, I thank Zvi Bern, David Kosower and my
other colleagues for many enjoyable collaborations on related topics.
The figures in this contribution were generated using 
Jaxodraw~\cite{Jaxo}, based on Axodraw~\cite{Axo}.
This work was supported by the US Department of Energy under contract
DE-AC02-76SF00515.


\end{document}